%% file: Interactions_FinalPaper_JINST.tex
\documentclass[cits]{JINST}

\usepackage{atlasphysics}

\usepackage{subfigure}
\usepackage{mathrsfs}
\usepackage{graphicx}
\usepackage{multirow}
\usepackage{rotating}
\usepackage{xspace}

\title{\vspace{-3.0cm}\flushleft{\normalsize\normalfont{CERN-PH-EP-2011-147}} \flushright{\vspace{-0.86cm}\normalsize\normalfont{Submitted to JINST}} \\[3.0cm] \flushleft{\bf A study of the material in the ATLAS inner detector using secondary hadronic interactions}}

\author{The ATLAS Collaboration}

\abstract{The ATLAS inner detector is used to reconstruct secondary vertices due to hadronic interactions of primary collision products, so probing the location and amount of  material in the inner region of ATLAS. Data collected in 7 \TeV \xspace \emph{pp} collisions at the LHC, with a minimum bias trigger, are used for comparisons with simulated events. The reconstructed secondary vertices have spatial resolutions ranging from $\sim 200~\mu$m to 1~mm. The overall material description in the simulation is validated to within an experimental uncertainty of about 7\%. This will lead to a better understanding of the reconstruction of various objects such as tracks, leptons, jets, and missing transverse momentum.}


\begin{document}

\section{Introduction}

An accurate description of material in the ATLAS inner detector is crucial to the understanding of tracking performance, as well as other reconstructed objects such as electrons, jets and missing transverse momentum. Traditionally, photon conversions, which are sensitive to the radiation length of material, are used to map the detector. To quantify the amount of material in terms of interaction lengths, the measurement must be converted from radiation lengths, which requires a very precise knowledge of the actual composition of the material, or a direct measurement of quantities sensitive to the interaction length must be made. This paper describes a direct measurement using the reconstruction of secondary vertices due to hadronic interactions of primary particles, and is based on a careful comparison of the secondary vertex yield in data with a simulation of the ATLAS inner detector.  The simulation implements precise information about the inner detector components, and this study aims to validate its correctness.

In addition to directly probing the number of hadronic interaction lengths of material, another advantage of studying such interactions is the excellent spatial resolution of the resulting reconstructed secondary vertices. This property is exploited to study the precise location of the material. Since hadronic interactions vertices usually result from low to medium energy primary hadrons (with average momentum, $<\!\! p \!\! >$ around 4 GeV, and with about 96\% having $p <10$ GeV) the outgoing particles have low energy and large opening angles between them. This contrasts with photon conversions, where the opening angle between the outgoing electron-positron pair is close to zero. Consequently, the technique presented here has a much improved spatial resolution. Hadronic interactions will often produce more than two outgoing particles with momenta high enough to be reconstructed by the tracking system. An inclusive vertex finding and fitting package is used to reconstruct these vertices.

This paper is structured as follows: Section~\ref{detector} gives a brief description of the inner detector, Section~\ref{samples} gives details of the data sample and track selection criteria used in this analysis, and Section~\ref{Algorithm} contains a description of the vertex-finding algorithm. Section~\ref{data_plots} contains qualitative results from data and comparisons with Monte Carlo simulations (MC). Section~\ref{systematics} describes the various systematic uncertainties, which are used in Section~\ref{compareSec} to make quantitative comparisons with MC.

\section{Inner Detector}
\label{detector}

\begin{figure}[htbp]
\begin{center}
\includegraphics[width= 1.0\textwidth]{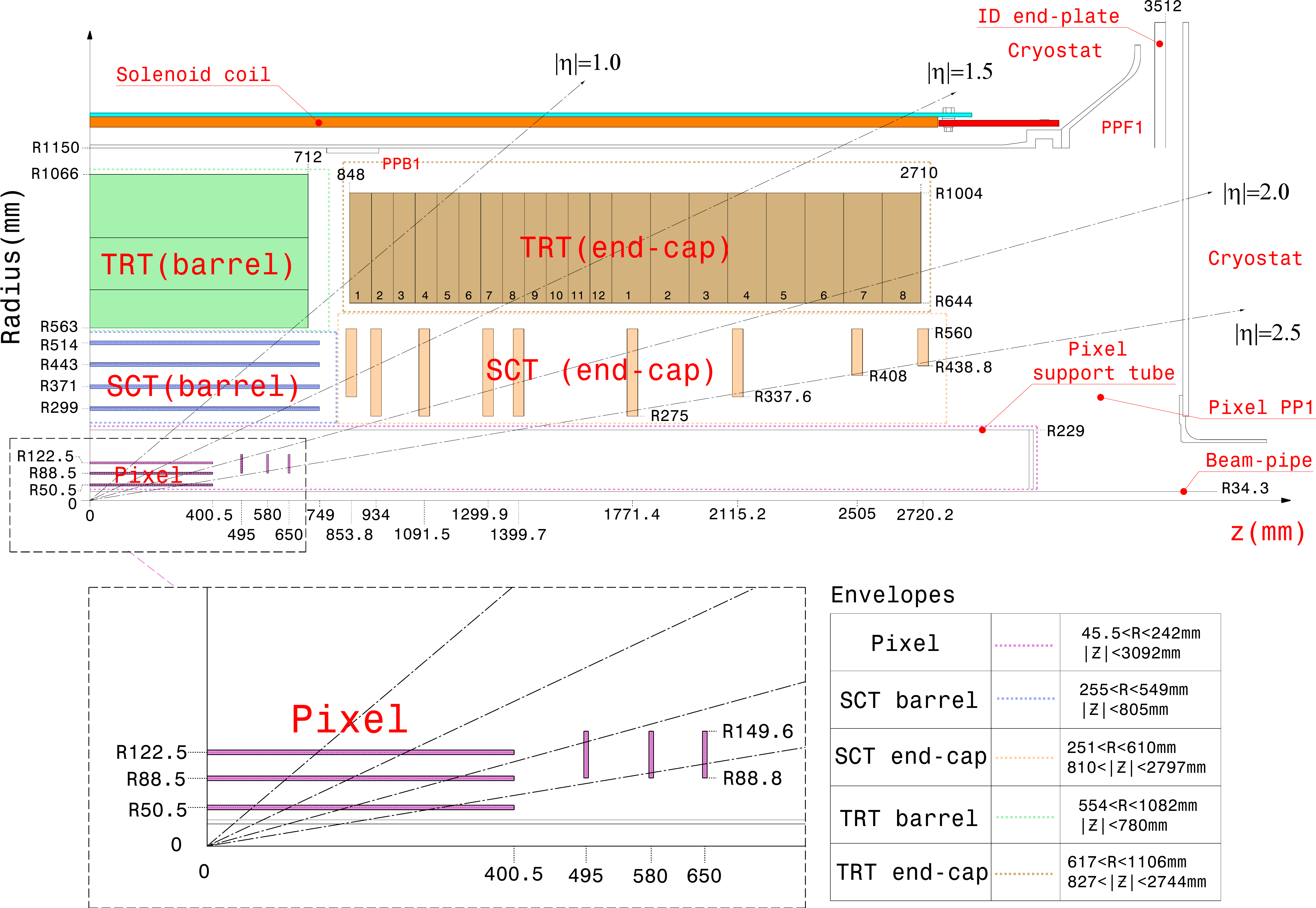}

\caption{Plan view of a quadrant of the inner detector showing each of the major
detector elements with their active dimensions and envelopes. The lower part shows a zoom of the pixel region.
\label{fig:indet}}
\end{center}
\end{figure}

The inner detector consists of a semi-conductor pixel detector, a semi-conductor microstrip detector (SCT), and a transition radiation tracker (TRT), all of which are surrounded by a solenoid magnet providing a 2 T field~\cite{detector1, detector2}. It extends from a radius\footnote{ATLAS uses a right-handed coordinate system with its origin at the nominal interaction point (IP) in the center of the detector and the $z$-axis along the beam pipe. The $x$-axis points from the IP to the center of the LHC ring, and the $y$ axis points upward. Cylindrical coordinates $(R,\phi)$ are used in the transverse plane, $\phi$ being the azimuthal angle around the beam pipe. The pseudorapidity is defined in terms of the polar angle $\theta$ as $\eta=-\ln\tan(\theta/2)$.} of about 45~mm to 1100~mm and out to $|z|$ of about 3100~mm. A quarter section of the inner detector is shown in Figure~\ref{fig:indet}. It provides excellent track impact parameter and momentum resolution over a large pseudorapidity range ($|\eta| < 2.5$), and determines the positions of primary and secondary vertices.

In the barrel region, the precision detectors (pixel and SCT) are arranged in cylindrical layers around the beam pipe, and in the endcaps they are assembled as disks and placed perpendicular to the beam axis. The TRT is made of drift tubes, which are parallel to the beam axis in the barrel region, and extend radially outward in the endcap region. The envelope of the barrel pixel detector covers the radial region from 45~mm to 242~mm, which includes the active layers, as well as supports, and extends to about $\pm 400$~mm in \emph{z}. The barrel SCT envelope ranges from 255~mm to 549~mm, and the barrel TRT sub-system covers the radial range from 554 to 1082~mm. The latter two sub-systems extend to about $\pm 800$~mm in \emph{z}. Outside the beam pipe, the regions without material are filled with different gases, N$_2$ and CO$_2$ in the silicon and TRT volumes, respectively, and for simplicity they are referred to as air gaps.

All pixel sensors in the pixel detector, in both barrel and end-cap regions, are identical and have a nominal size of $50\times400$~$\mu$m$^2$, and there are approximately 80.4 million readout channels. The pixel detector in the barrel region has three layers, containing 22, 38, and 52 staves in azimuth, respectively. The layers are concentric with the beam pipe. Each stave contains 13 modules along \emph{z}, and each module contains about 47000 individual pixels. A `zoomed-in' view of a module can be seen in Figure~4.4 of Ref.~\cite{detector2}. In the SCT barrel region there are small angle stereo strips in each layer, with one set parallel to the beam direction to measure $R-\phi$ and the other set at an angle of 40 mrad, which allows for a measurement of the \emph{z}-coordinate. The endcap SCT detector has a set of strips running radially outward and a set of stereo strips at an angle of 40 mrad to the former. The total number of readout channels in the SCT is approximately 6.3 million. The TRT consists of 298,000 drift tubes with diameter 4~mm, and provides coverage over $|\eta| < 2.0$. The material measurements in this paper are focused mainly on the beam pipe and the pixel detector in the barrel region.

\section{Data samples, track selection and reconstruction}
\label{samples}

\subsection{Data samples}
\label{data_samples}

The data used in this analysis were collected during March-June 2010 in proton-proton collisions at a center-of-mass energy of 7 \TeV. During this initial period the instantaneous luminosity was approximately
$10^{27}-10^{29}$~cm$^{-2}$ s$^{-1}$. Data were collected using minimum bias triggers~\cite{Syst1} and correspond to approximately 19 nb$^{-1}$ of integrated luminosity. Later runs with higher instantaneous luminosity were not used. The minimum bias triggers collect single-, double- and non-diffractive events, with the majority belonging to the last category. In order to facilitate comparisons with MC, single- and double-diffractive contributions are effectively removed from the data and the remaining events are compared with a simulated sample of non-diffractive events~\cite{Syst1}. This is achieved by requiring a large track multiplicity at the primary vertex. This approach works best when the number of additional \emph{pp} interactions per event (pile-up) is small, and so only the low luminosity runs are used. It is required that there be exactly one reconstructed primary vertex in the event, and that it should have at least 11 associated tracks; this requirement is expected to keep less than 1\% of single- and double-diffractive events, while retaining $\sim 68\%$ of non-diffractive events. At this stage there are $\sim$ 40.9 (13.5) million events in data (MC), respectively. MC events are weighted such that the mean and width of the \emph{z}-coordinate distribution of the primary vertex position match the data. MC events were generated using {\sc pythia6}~\cite{pythia6} with the {\sc ambt1} tune~\cite{ambt1}, simulated with {\sc geant4}~\cite{geantref}, and processed with the same reconstruction software as data. The ATLAS simulation infrastructure is described elsewhere~\cite{atlasSimu}.

\subsection{Track selection}
\label{trk_samples}

Since the main goal of the track reconstruction software is to find particles originating from the primary vertex, it puts stringent limits on the allowed values of transverse and longitudinal impact parameters. As a result, the reconstruction efficiency for secondary track candidates strongly depends on both \emph{R}- and \emph{z}-coordinates of the vertex they originate from.

In order to reconstruct secondary interactions, well-measured secondary track candidates should be selected, and tracks coming from the primary vertex rejected in order to reduce combinatorial background. Tracks are required to have: (a) transverse momentum above 0.3~\GeV, (b) transverse impact parameter relative to the primary vertex of at least 5~mm, and (c) fit $\chi^2$/dof $< 5$. The impact parameter requirement removes more than 99\% of the primary tracks, as well as many tracks produced in $\kshort$ decays and $\gam$ conversions. In general, particles produced in secondary hadronic interactions have much larger impact parameters, especially in comparison to $\gam$ conversions, which tend to point back to the primary vertex.  There is no requirement on the number of hits in the pixel detector, since that would limit the scope of this analysis to a radius less than that of the third layer. However, tracks are required to have at least one hit in the SCT. Constraints on the track reconstruction are such that the efficiency to find tracks arising from secondary vertices with $|z|>300$~mm is very low, consequently this region is not considered when making quantitative comparisons of the rate of vertex yields per event in data and MC.

\subsection{Track reconstruction in data and MC}
\label{trackMC}

Extensive studies of track reconstruction algorithms have been performed in data and MC and generally the data are found to be well simulated by the MC, but there is some disagreement in the number of reconstructed primary tracks~\cite{Syst1}. This has a cascade effect on the analysis, in that having more primary particles in data implies that there will be more secondary interactions, leading to more secondary particles that can further interact in outer layers. Hence, when comparing the number of reconstructed secondary vertices per event in data and MC, the raw yield in MC is multiplied by a correction factor. To determine this correction, all reconstructed primary tracks are extrapolated to find their intersections with inner detector material layers, and only tracks which intersect a layer with $|z|<300$~mm are considered further. To account for the fact that primary tracks produced at small polar angles travel through more material thus resulting in a higher interaction probability, each track is weighted by $1/\sin\theta$, where $\theta$ is its polar angle. The ratio of the weighted sum of the number of tracks (in data and MC) gives average correction factors, which are estimated to be 1.072 at the beam pipe, 1.071, 1.061, and 1.059 at the first, second and third pixel detector layers, respectively, and 1.057 at the first SCT layer. These scaling factors are also used when comparing various distributions in data and MC.

\begin{figure}[htbp]
\begin{center}
\includegraphics[width= 1.0\textwidth]{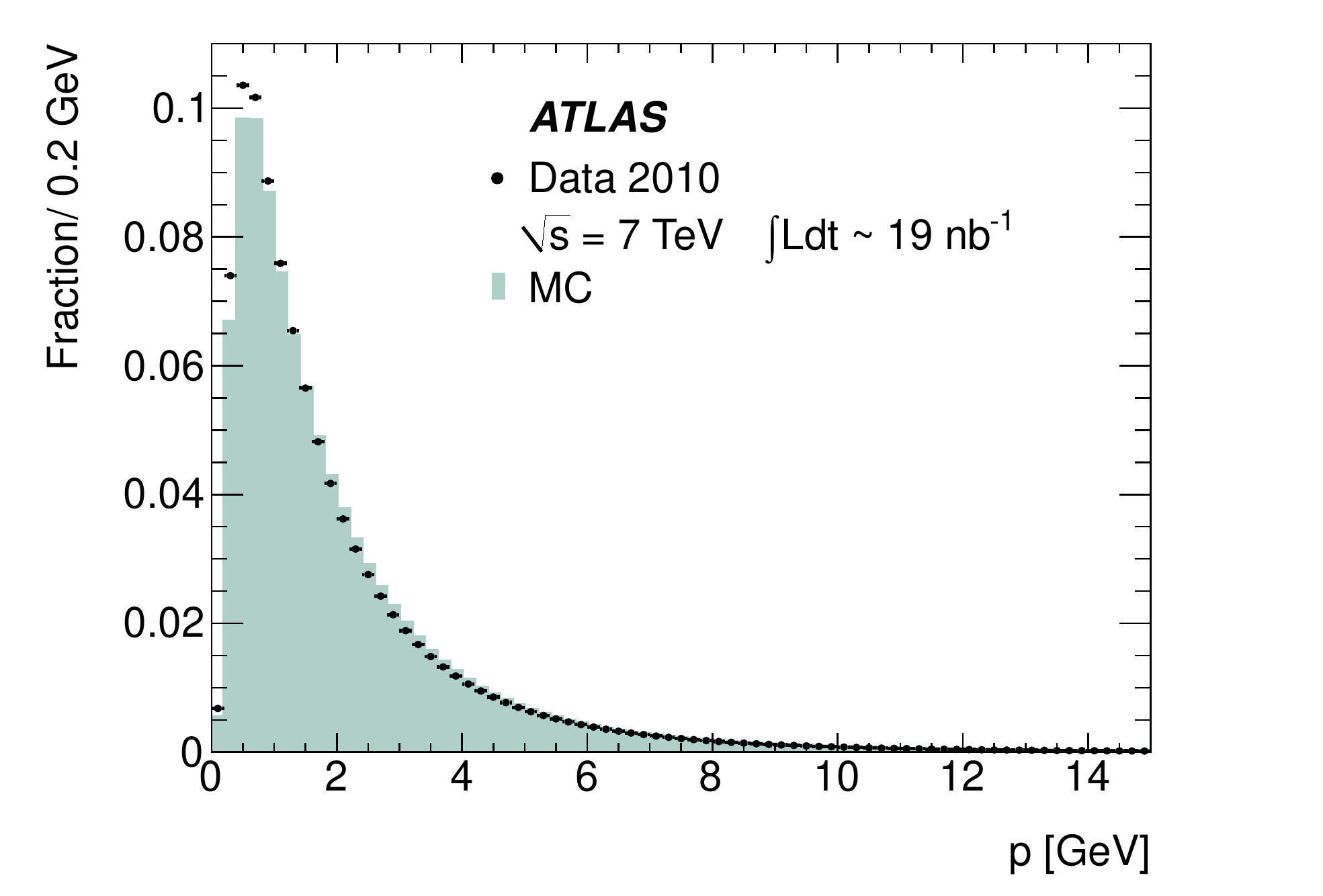}

\caption{Momentum spectrum of primary tracks intersecting the beam pipe at $|z| < 300$~mm in data (points) and MC (filled histogram). The two spectra are normalized to unity.
\label{fig:compareDataMCPrimTrkBP}}
\end{center}
\end{figure}

The momentum spectra of primary tracks in data and MC agree reasonably well. Figure~\ref{fig:compareDataMCPrimTrkBP} shows the momentum spectra of primary tracks that intersect the beam pipe with $|z|<300$~mm, after weighting as described above. Differences in the momentum spectrum could also lead to mismodelling of interactions. This is checked by reweighting the MC momentum spectrum to match the data, and no significant effect was observed.

\section{Description of vertex reconstruction and resolution}
\label{Algorithm}

\subsection{Vertex reconstruction}

A \emph{pp} collision event may have decays of short-lived particles, $\kshort$ and $\Lambda$ decays, $\gam$ conversions, and several material interaction vertices with \emph{a priori} unknown multiplicity. Clean detection of material nuclear interactions requires reconstruction and elimination of all other secondary vertices. A universal vertex finder, designed to find all vertices in the event, is used in this analysis.

The algorithm starts by finding all possible intersections of pairs of selected tracks. It assumes that these two secondary tracks are coming from a single point and determines the vertex position and modifies track parameters to satisfy this assumption. Differences between the measured track parameters and the recalculated ones define the vertex $\chi^2$. The reconstructed two-track vertices define the full vertex structure in the track set because any \emph{N}-track vertex is simply a union of corresponding two-track sub-vertices. Requiring these vertices to have an acceptable $\chi^2$ ($<4.5$) removes $\sim 85\%$ of random pairings, and MC studies indicate that more than 83\% of nuclear interaction vertices are retained. To further reduce the number of fake vertices from random combinatorics, tracks must not have hits in silicon layers at a radius smaller than the radius of the reconstructed vertex, and must have hits in some layers that are at larger radii than the vertex. Vertices that fail this criterion are removed from the list of selected two-track vertices. According to MC, this procedure removes, depending on radius, anywhere from half to two-thirds of the initial set of two-track vertices, with only a 2-10\% reduction in efficiency for reconstructing nuclear interaction vertices.

To finalize vertex finding, the total number of vertices in the event is minimized by merging the two-track candidates that are nearby; this decision is based on the separation between vertices combined with the vertex covariance matrices. Initially, any track can be used in several two-track vertices. Such cases also must be identified and resolved so that all track-vertex associations are unique. The algorithm performs an iterative process of cleaning the vertex set, based on an incompatibility-graph approach~\cite{Graph}. At each step it either identifies two close vertices and merges them, or finds the worst track-vertex association for multiply assigned tracks and breaks it. Iterations continue until no close vertices or multiply-assigned tracks are left. This algorithm successfully works on events with track multiplicity up to $\sim$200, which is significantly larger than the average multiplicity in events used in this analysis ($\sim 50$ tracks/event).

\subsection{Vertex resolutions}
\label{vtxres}

\begin{figure}[htbp]
\centering
\mbox{\subfigure[]{\includegraphics[width=0.5\textwidth]{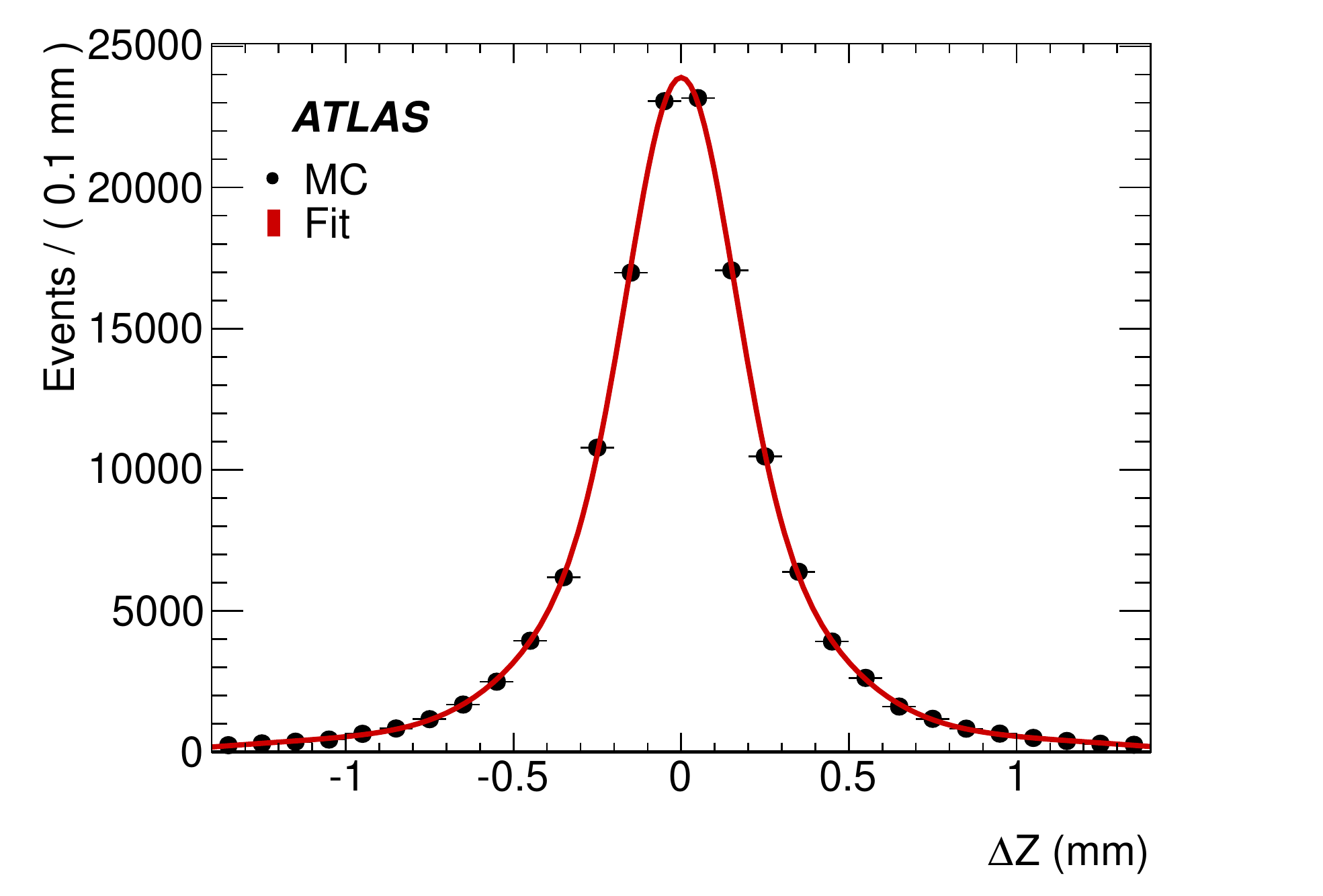}}\quad
\subfigure[]{\includegraphics[width=0.5\textwidth]{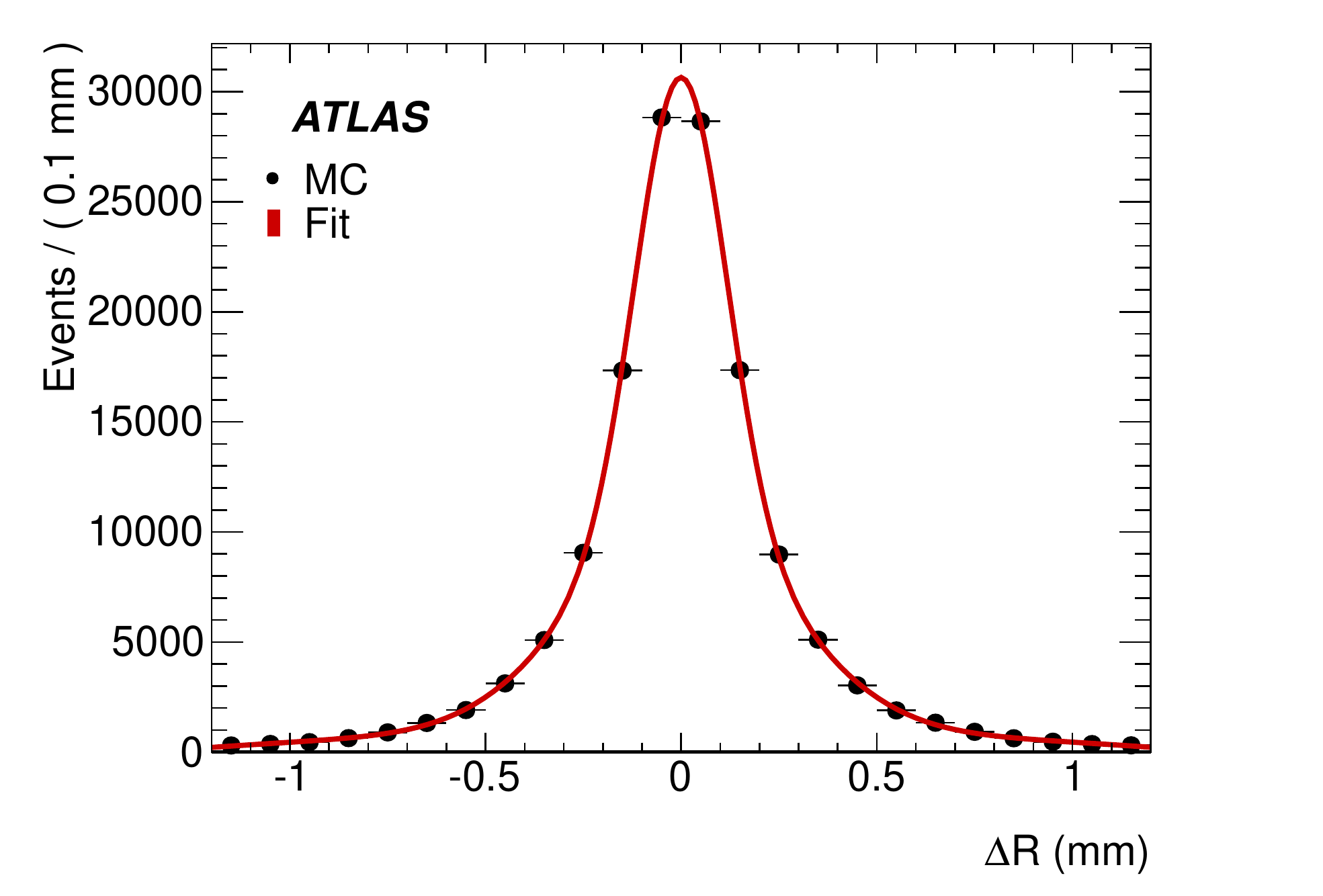}}}
\caption{Non-diffractive MC: (a) \emph{z} and (b) \emph{R} resolutions for vertices reconstructed in the beam pipe.
\label{fig:resPlots}}
\end{figure}

The spatial resolution of the reconstructed vertices depends on the quality of track reconstruction. MC studies indicate that the resolution for hadronic interaction vertices is 200-300~$\mu$m (in both \emph{R} and \emph{z}) for reconstructed vertices with $R \leq 100$~mm and $\sim 1$~mm for vertices at larger radii. The resolution along the $\phi$ coordinate, i.e., transverse to \emph{R}, is 100-140~$\mu$m, depending on the radius of the vertex. At smaller radius, tracks have more hits in the pixel and SCT detectors, so their parameters are better determined. In contrast, the radial resolution in photon conversions is approximately 5 mm~\cite{detector1}. Figure~\ref{fig:resPlots} presents the \emph{z} (left) and \emph{R} (right) resolutions for nuclear interaction vertices reconstructed at the beam pipe, where the signal is fitted with a sum of two Gaussian functions (with a common mean), and the background is represented by a first-order polynomial. The width of the core and the fraction of entries in it varies with radius: at the beam pipe they range from $120~\mu$m to $150~\mu$m, and $50\%$ to 54\%, for \emph{R} and \emph{z}, respectively. The corresponding numbers for the $\phi$-coordinate are $60~\mu$m, and 65\%, respectively. Resolutions for vertices with more than two tracks are slightly better than for vertices with only two tracks. For instance, at the beam pipe $\sim 96\%$ of the two-track vertices are within $\Delta R < 1$~mm, whereas for vertices with more than two tracks $\sim 95\%$ have $\Delta R < 0.6$~mm, where $\Delta R$ is the difference between the radii of the true nuclear interaction and reconstructed vertex positions.

\section{Reconstructed vertices in 7 \TeV \xspace data}
\label{data_plots}

\begin{figure}[htbp]
\begin{center}
\includegraphics[width= 1.0\textwidth]{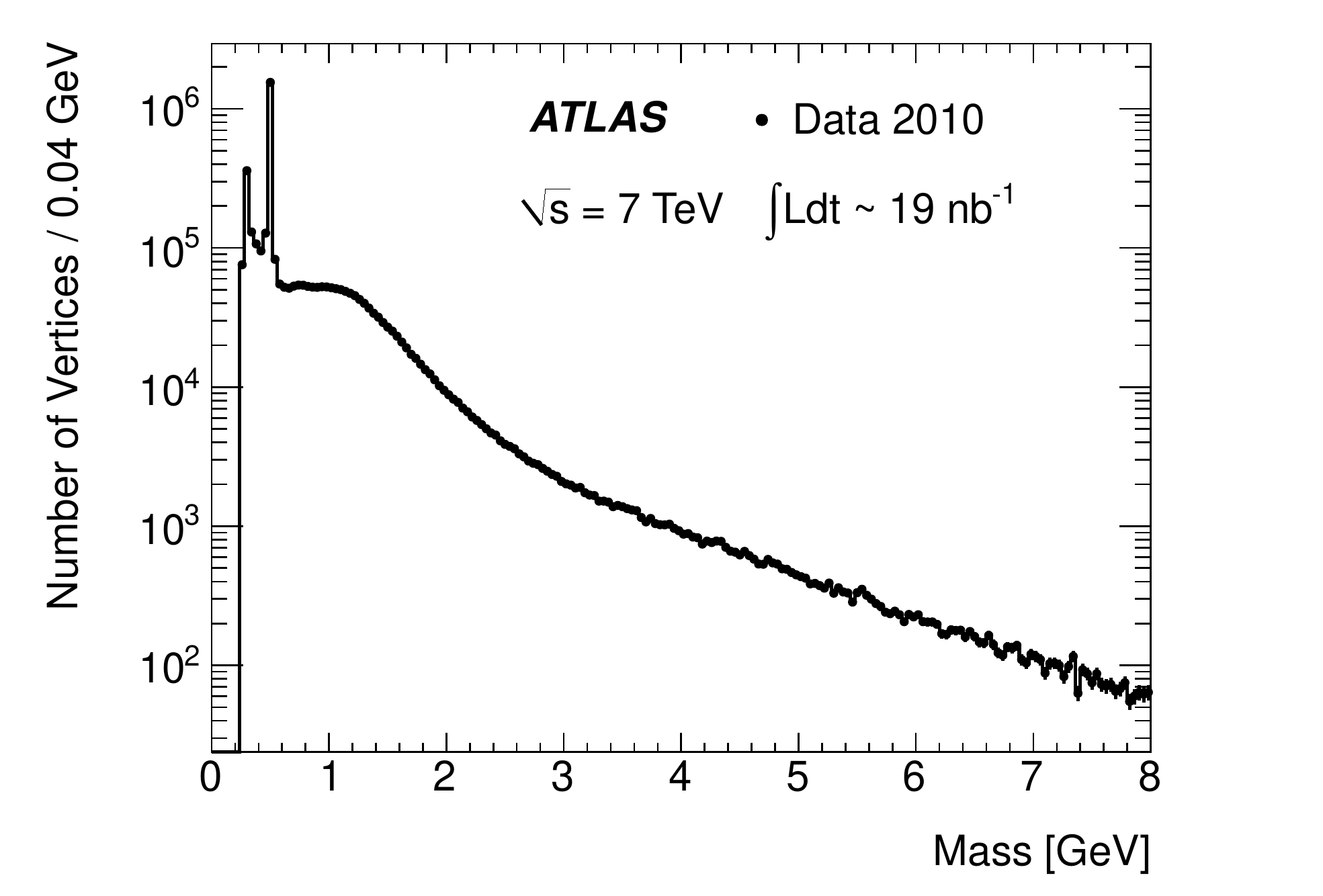}

\caption{Mass of reconstructed vertices in data. All secondary vertices with $|z|<700$~mm have been used.
\label{fig:dataMass}}
\end{center}
\end{figure}

When removing fake two-track vertices, as described previously, no attempt is made to remove $\gam$, $\kshort$, $\Lambda$ candidates; these are vetoed at a later stage. The distribution of the reconstructed invariant mass of charged particles associated to each secondary vertex is shown in Figure~\ref{fig:dataMass}, assuming the pion mass for each track. A clear $\kshort$ peak can be seen, as well as the smaller peak at threshold due to $\gam$ conversions. Their mass is not zero because pion masses are incorrectly attributed to the electrons. The 5~mm minimum requirement on the transverse impact parameter of tracks has already strongly suppressed conversions. The `shoulder' at $\sim$ 1200 \MeV \xspace is a kinematic effect and reflects the minimum requirement on track $\pt$ (its position changes with this threshold value). The remaining $\gam$ conversion candidates are vetoed by removing vertices with an invariant mass less than 310 \MeV. Similarly, $\kshort$ candidates are removed if the invariant mass lies within $\pm 35$ \MeV \xspace of the nominal $\kshort$ mass, and $\Lambda$ candidates\footnote{For the $\Lambda$ veto, the track with the larger momentum is assumed to be the proton. According to MC, for $\sim$~3\% of $\Lambda$'s the proton has the lower momentum. Since the number of reconstructed interaction vertices is about 15 times the number of reconstructed $\Lambda$'s, many of which decay in the gaps between material layers, any resulting contamination is small and is neglected.} are vetoed if the mass lies within $\pm 15$~\MeV \xspace of the nominal $\Lambda$ mass. These mass vetoes have been applied to all the following figures and results.

\begin{figure}[htbp]
\begin{center}
\includegraphics[width= 1.0\textwidth]{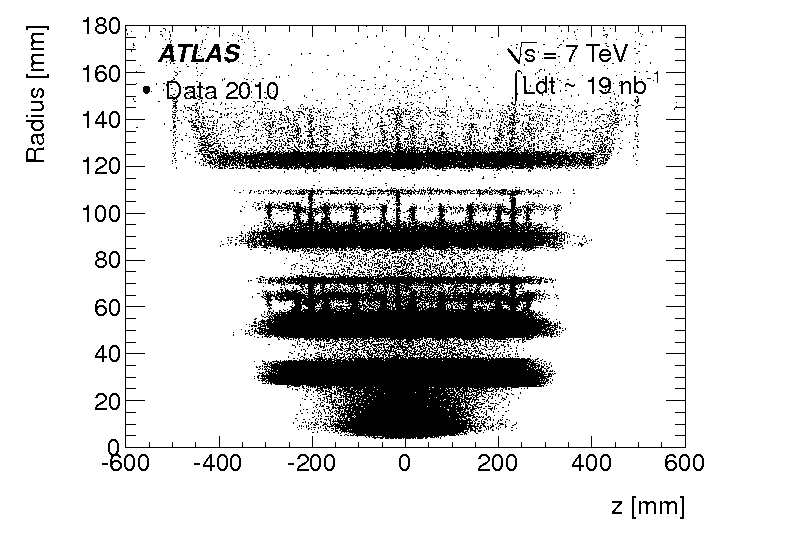}

\caption{The \emph{R} vs. \emph{z} distribution of secondary vertices reconstructed in data. The bin width is 7 mm in \emph{z} and 1 mm in \emph{R}. To aid the eye, only bins with 5 or more entries have been displayed.
\label{fig:dataRZ}}
\end{center}
\end{figure}

\begin{figure}[htbp]
\begin{center}
\includegraphics[width= 1.0\textwidth]{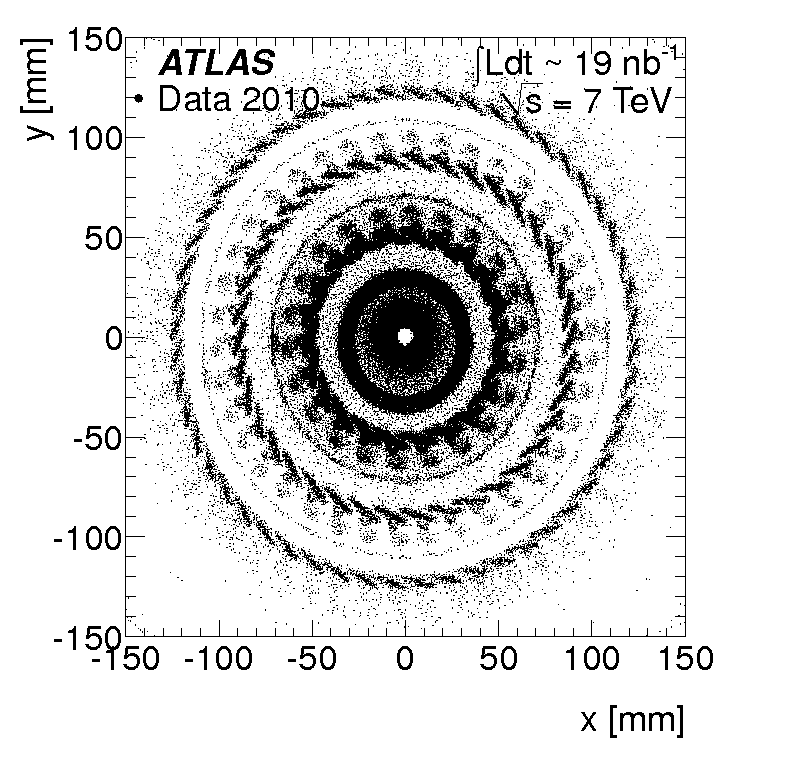}

\caption{The \emph{y} vs. \emph{x} distribution of secondary vertices reconstructed in data. The bin width is 1.5 mm along both axes. All secondary vertices with $|z|<700$~mm have been used. To aid the eye, only bins with 5 or more entries have been displayed.
\label{fig:dataXY}}
\end{center}
\end{figure}

Figure~\ref{fig:dataRZ} shows the \emph{R} vs. \emph{z} distribution of the secondary vertices. MC studies indicate that vertices inside the beam pipe and almost all of the vertices in the gaps between material surfaces are due to combinatorial background, with a very small fraction of the latter due to interactions with the gases in these gaps (the density of silicon is about 1000 (1500) times the density of CO$_2$ (N$_2$)). The beam pipe envelope, consisting of a Beryllium cylinder, followed by layers of aerogel, kapton tape and coatings, extends from a radius of 28~mm to 36~mm; studies of this region are described later in the paper. The horizontal bands at R$\sim$ 47, 85, 120~mm include the pixel detector modules and the bands at R$\sim$ 65, 70, 105, 110~mm represent cables, services and supports.  The pixel modules and their support staves are tilted by 20$^{\circ}$ in the \emph{xy} plane and by 1.1$^{\circ}$ with respect to the beam axis. Furthermore, the pixel stave includes a $\sim$ 2~mm (radius) cooling pipe and carbon supports of varying thickness ($\le$ 2mm).  All these factors contribute to the visual thickness of the pixel layers. Details of the pixel module structure are discussed later. The vertical bands at various \emph{z} values are supports. Figure~\ref{fig:dataXY} presents the \emph{y} vs. \emph{x} position of vertices. The beam pipe and the three layers of the pixel detector are clearly visible. The $\phi$ structure of the pixel detector can also be seen.

\subsection{Qualitative comparison of data and MC}
\label{qualitative}

\begin{figure}[htbp]
\begin{center}
\includegraphics[width= 1.0\textwidth]{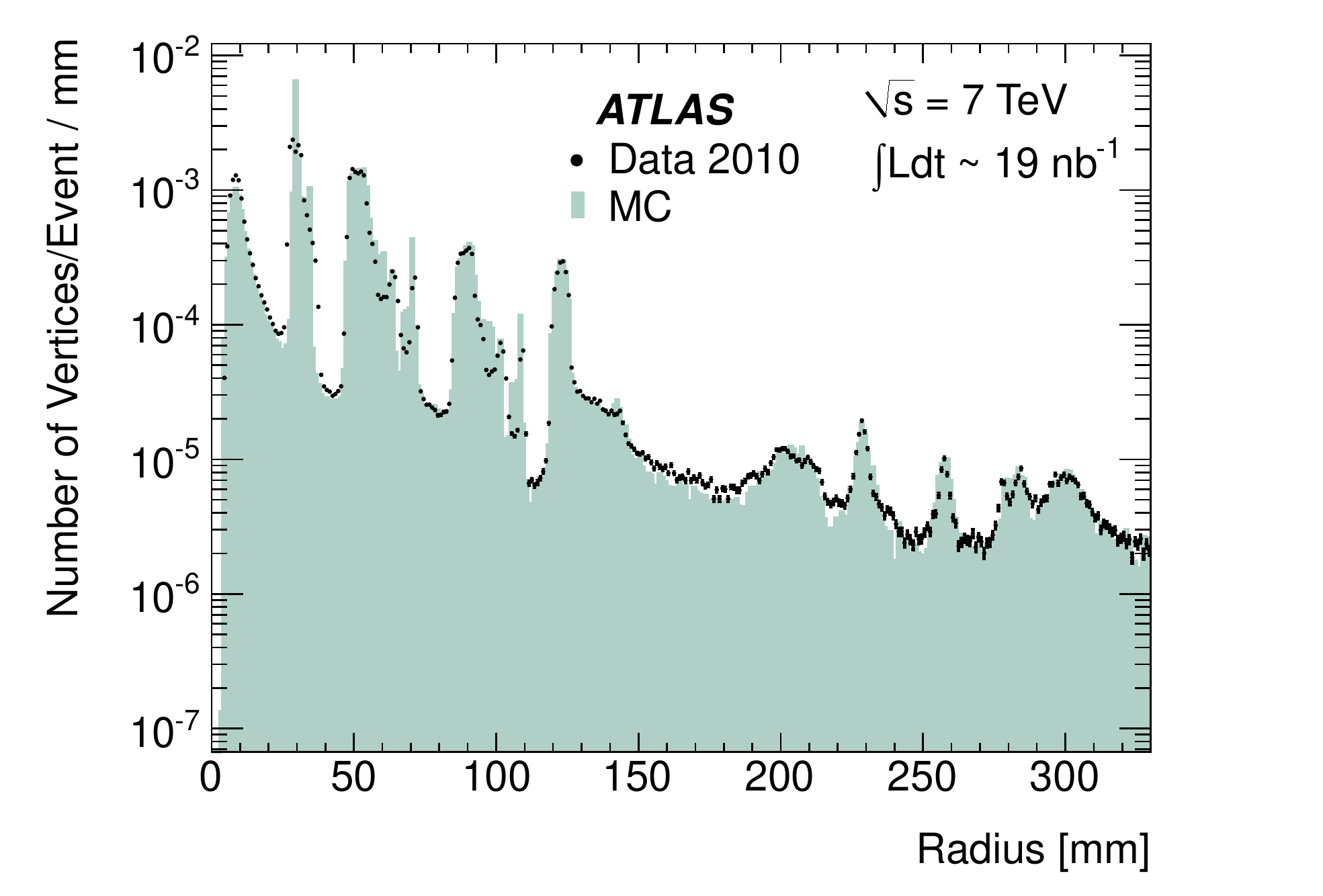}

\caption{Radial positions of reconstructed vertices for data (points) and MC (filled histogram), with $|z|<300$~mm.
\label{fig:dataMCcompareR}}
\end{center}
\end{figure}

\begin{figure}[htbp]
\begin{center}
\includegraphics[width= 1.0\textwidth]{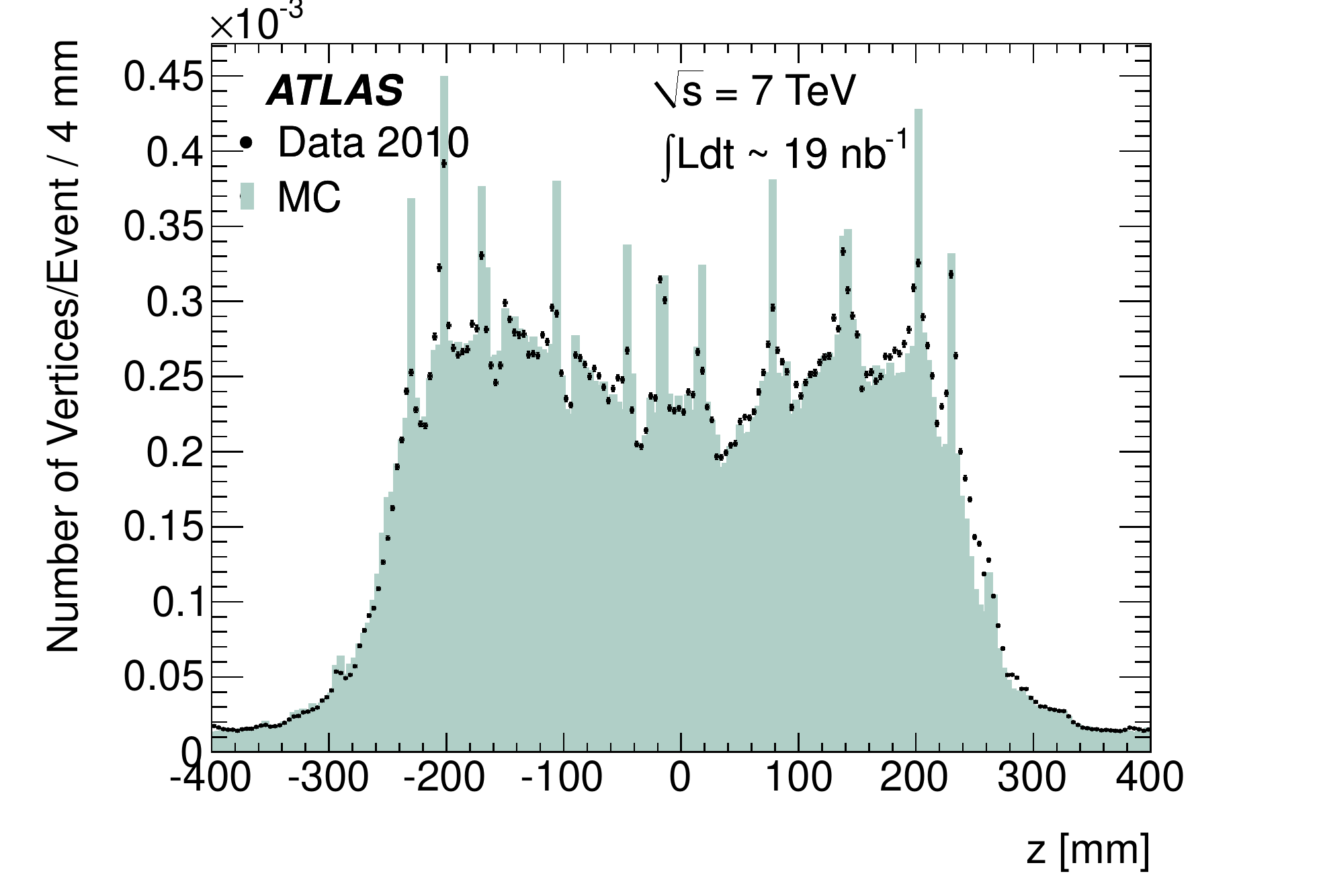}

\caption{\emph{z} positions of reconstructed vertices for data (points) and MC (filled histogram) for radius at or outside the beam pipe.
\label{fig:dataMCcompareZ}}
\end{center}
\end{figure}

Figures~\ref{fig:dataMCcompareR} and~\ref{fig:dataMCcompareZ} compare the  data and MC distributions of the \emph{R} and \emph{z} positions of the reconstructed vertices. For the \emph{z} projection, the reconstructed vertex radius is required to be at least that of the beam pipe, and for the radial projection, the vertex must have $|z| <300$~mm. In general, agreement in both shape and absolute rate is very good.  The fraction of vertices inside the beam pipe, comprised of decays of \emph{b}- and \emph{c}-hadrons, $\kshort$, strange baryons, and random combinatorial background due to primary tracks that fulfill the track selection criteria, is slightly larger in data than in MC (21\% vs. 19\%, respectively). The data also have slightly more entries in the air gaps between material layers. This is not unexpected as most of the vertices reconstructed in these regions are combinatorial background due to various track categories, and the simulation, although very good, does not make perfect predictions. Differences in the radial distribution at material layers are discussed in the next section. The \emph{z} positions of the reconstructed vertices in data agree well with MC, but there are some differences, e.g., in the region $200< |z| <300$~mm. Also, the `spikes' are sharper in the latter.

\begin{figure}[htbp]

    \begin{center}

        \subfigure[\emph{z} at the beam pipe]{%
           \includegraphics[width=0.5\textwidth]{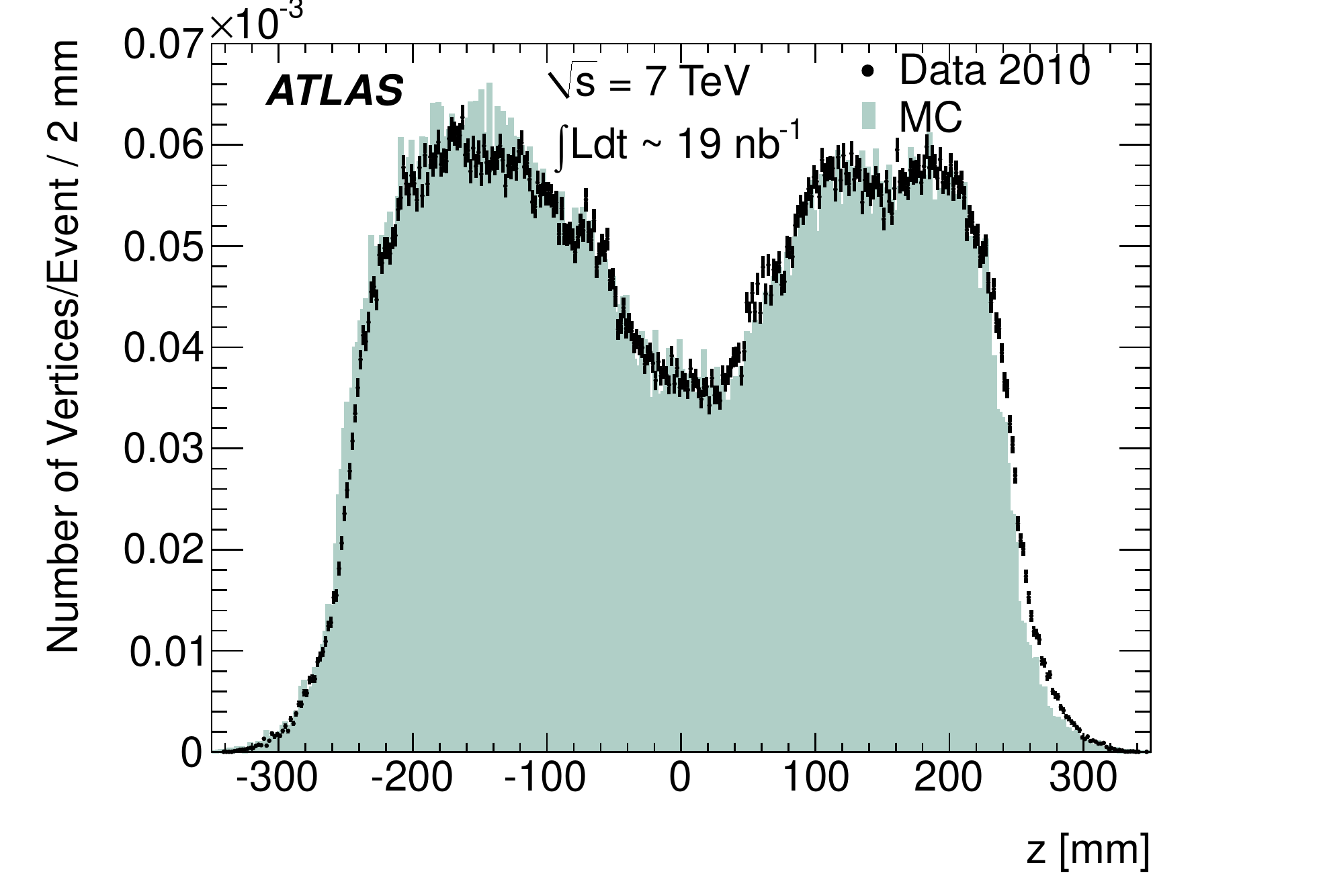}
        }%
        \subfigure[\emph{z} at the first pixel detector layer]{%
           \includegraphics[width=0.5\textwidth]{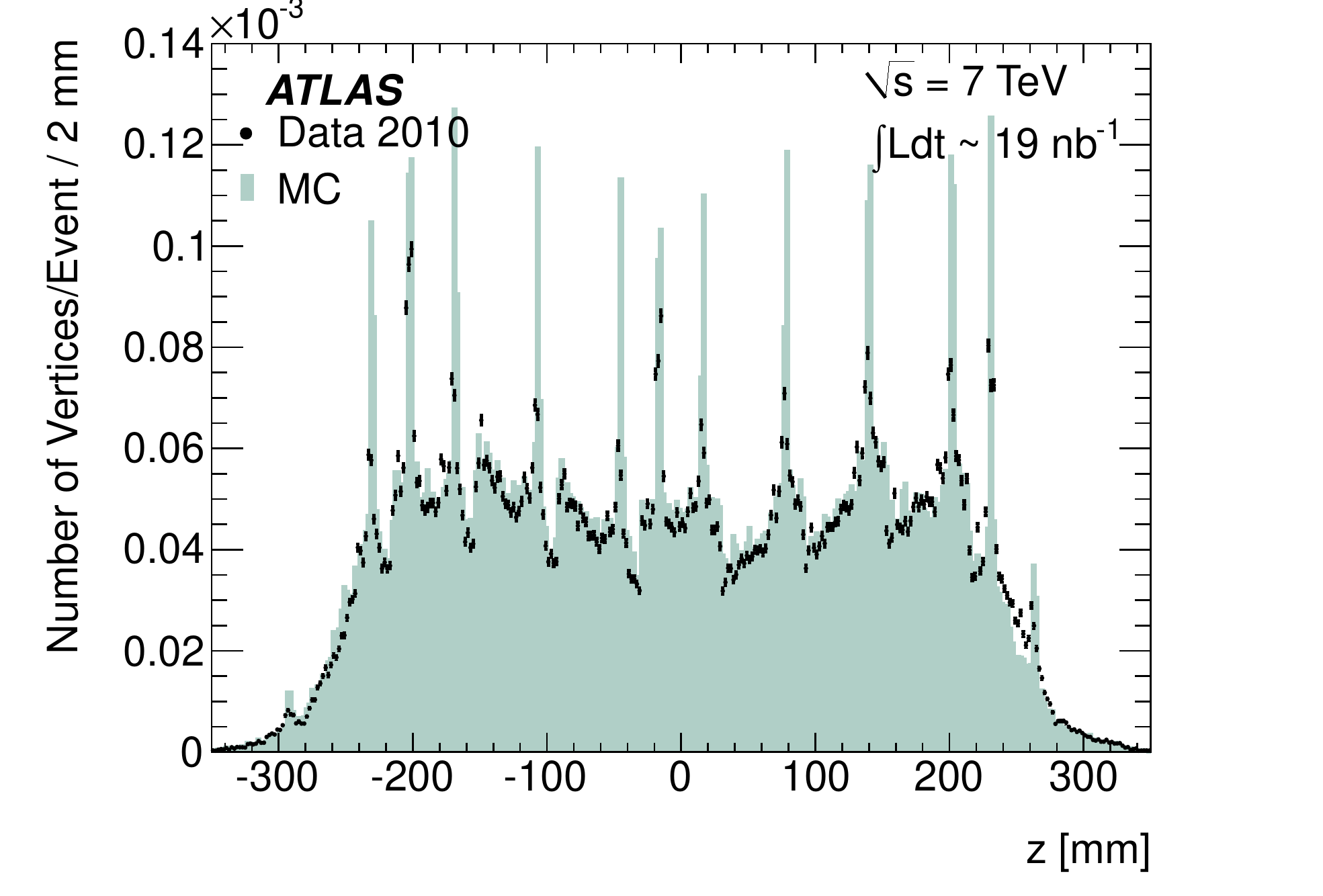}
        }\\ %
        \subfigure[\emph{z} at the second pixel detector layer]{%
            \includegraphics[width=0.5\textwidth]{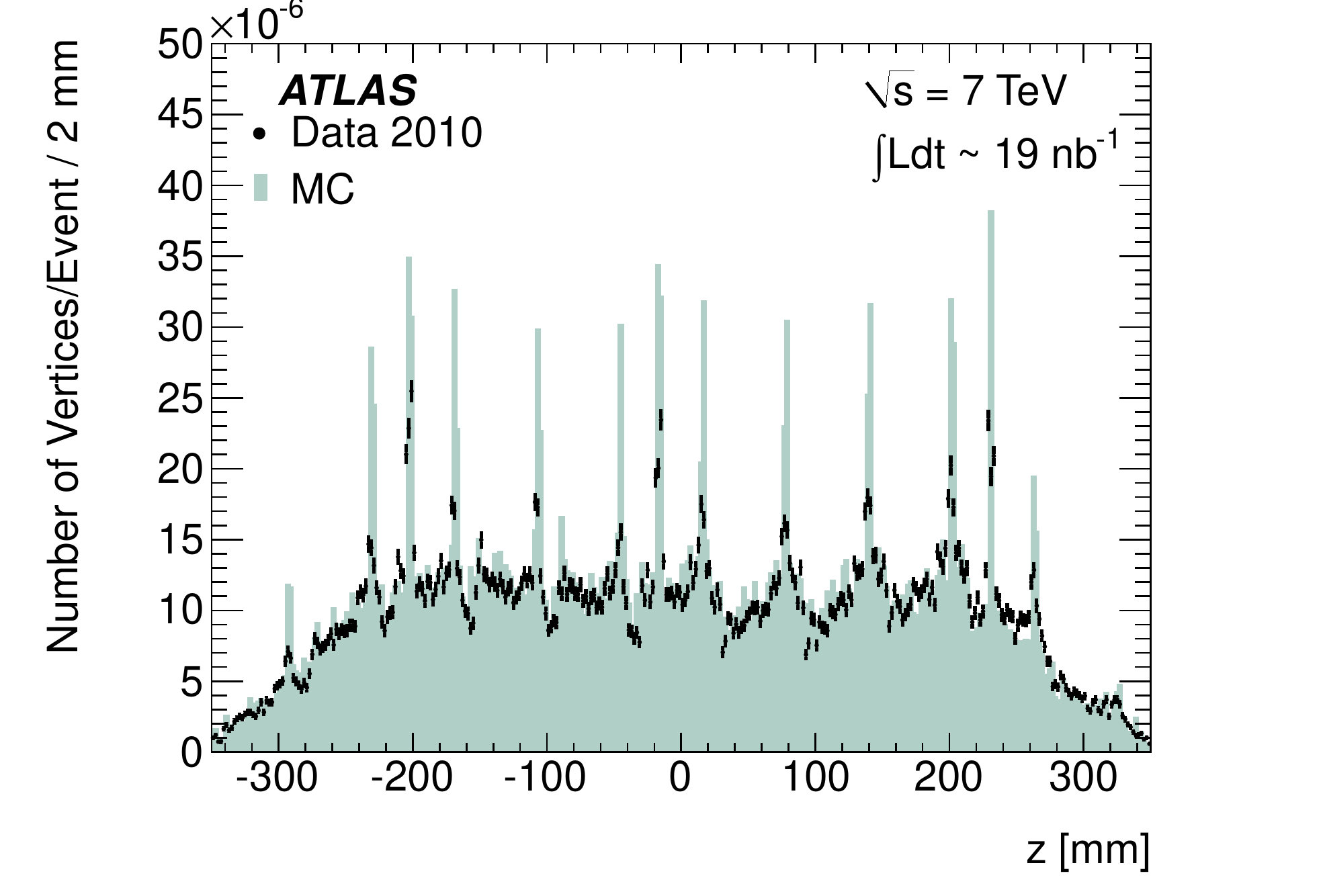}
        }%
        \subfigure[\emph{z} at the third pixel detector layer]{%
            \includegraphics[width=0.5\textwidth]{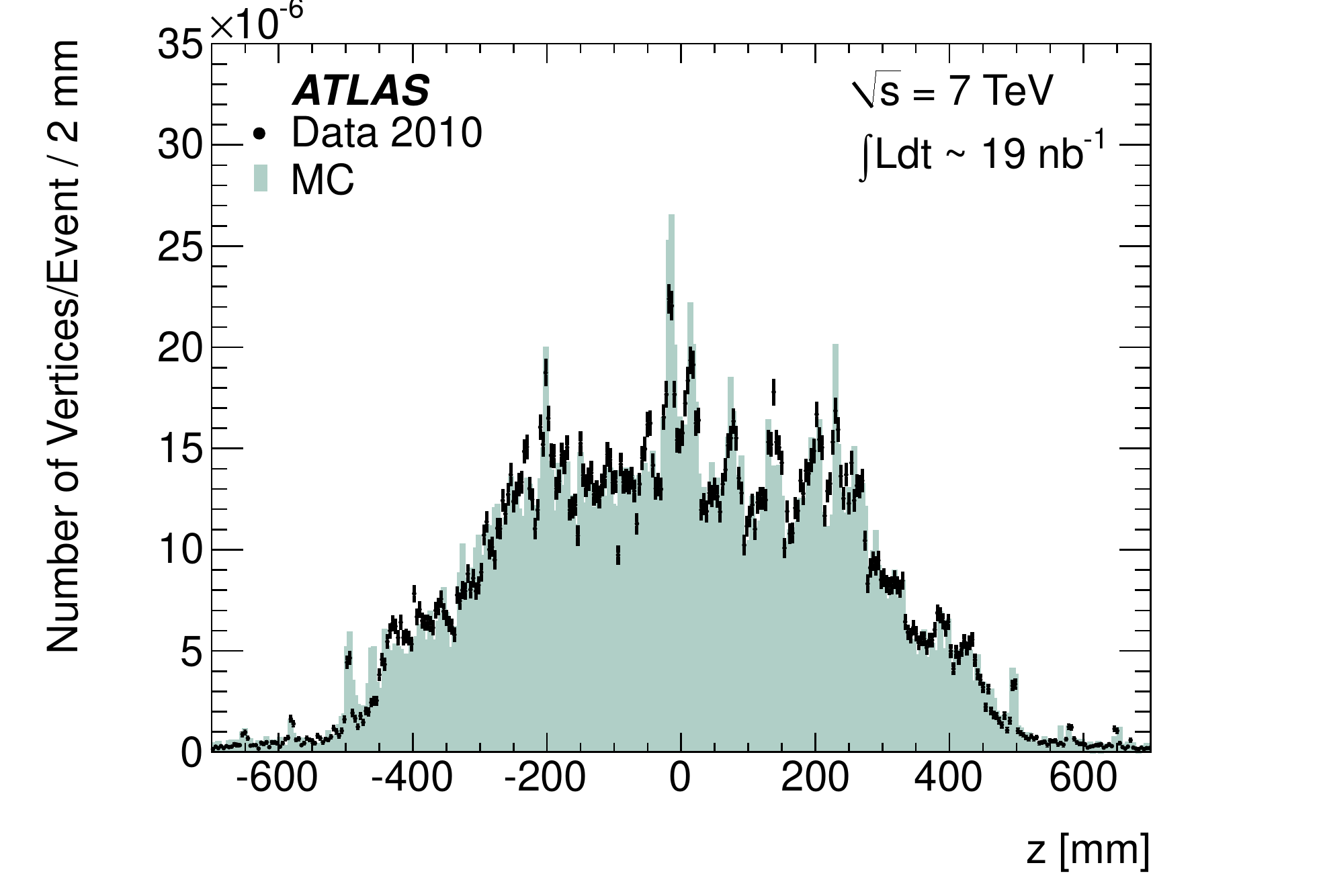}
        }%

        \caption{\emph{z} positions of reconstructed vertices for data (points) and MC (filled histogram). The radial ranges used for these projections are, (a) 29-35~mm, (b) 46.5-73.5~mm, (c) 85-111~mm, and (d) 118-143~mm. \label{fig:ZbpPix123}}
     \end{center}
\end{figure}

To study the \emph{z} position distributions in more detail, various material layers are considered individually. The \emph{z} projections at the beam pipe and the pixel detector are shown in Figure~\ref{fig:ZbpPix123}. The radial values used to make the \emph{z }projections were chosen after correcting for the position offsets of the beam pipe and the pixel detector, i.e., the radius of the vertex is calculated relative to the actual center rather than (0,0). This procedure is discussed in the next section. It is clear from Figure~\ref{fig:ZbpPix123}~(a), that MC underestimates the data at $200< z <300$~mm and overestimates at $-300 < z < -200$~mm. Note that the mean \emph{z} position of the primary vertex in this MC sample is at --5~mm, whereas in data it is much closer to 0. During track reconstruction a symmetric cut is placed on the maximum allowed value of the \emph{z} position of a track's point of closest approach to the beam axis, which is required to be within $\pm 250$~mm. This causes a small \emph{z} asymmetry in the number of tracks, which propagates to the \emph{z} positions of secondary vertices. This effect is diluted as the secondary vertex radius increases, but can still be seen in the corresponding distributions for the first two layers of the pixel detector. In the third layer, there appears to be a mismatch at $-470 <z < -400$~mm. Additionally, most of the `spikes' are much sharper in the MC than in data. This is partly because the MC has a simplified geometry for some detector elements, and partly because misalignments in data can cause broadening.

\subsection{Position of beam pipe and pixel detector}
\label{beampipePos}

\begin{figure}[htbp]
\centering
\mbox{\subfigure[$\phi$ vs. \emph{R}]{\includegraphics[width=0.5\textwidth]{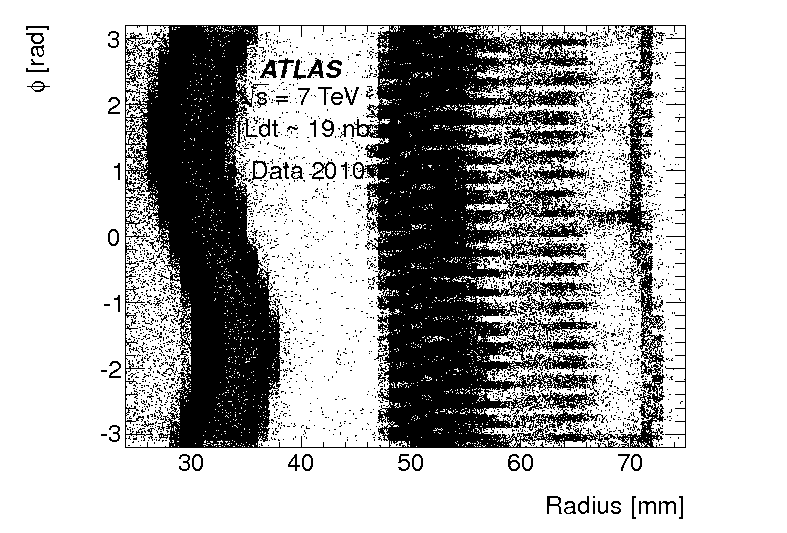}}\quad
\subfigure[$\phi$ profile at beam pipe]{\includegraphics[width=0.5\textwidth]{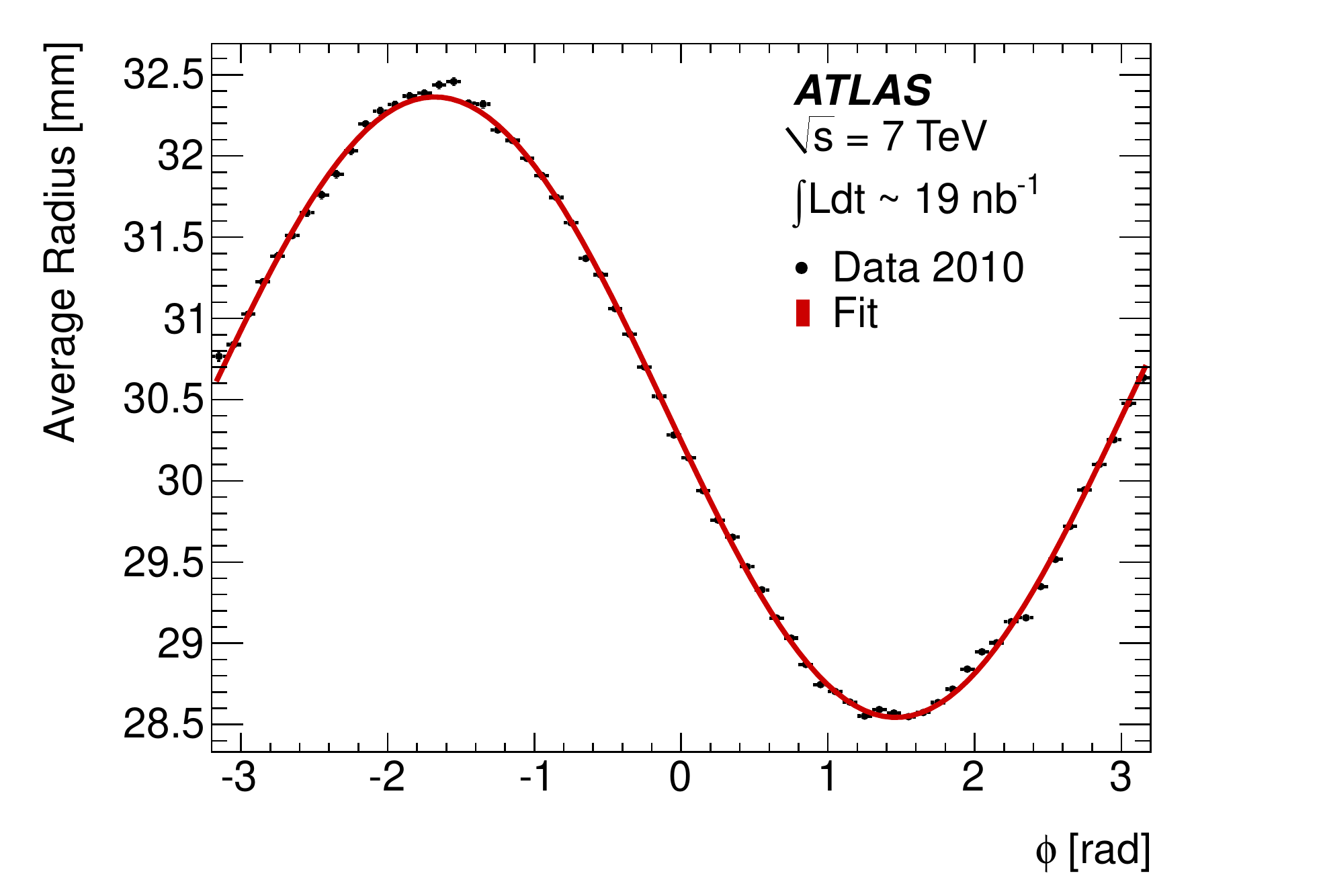}}}

\caption{(a) $\phi$ vs. \emph{R} of reconstructed vertices in the beam pipe and the first layer of the pixel detector, with $|z|<300$~mm. The bin width is 1~mm in radius, and 0.1 in $\phi$. To aid the eye, only bins with 20 or more entries are displayed. (b) A fit to the $\phi$ profile of reconstructed vertices. The \emph{y}-axis is the mean radius in a small range around the beam pipe ($26\leq R<39$~mm) for each $\phi$ bin. The bin width is 0.1 in $\phi$.
\label{fig:dataPhiBP}}
\end{figure}

In Figure~\ref{fig:dataMCcompareR}, the beam pipe appears to be broader in data than in MC. In reality it is not centered around (0,0). This is clearer in Figure~\ref{fig:dataPhiBP}~(a), which shows the $\phi$ vs. \emph{R} coordinates of the found vertices in data at the beam pipe and the first layer of the pixel detector. The sinusoidal behavior is a signature of an object not being centered around the nominal origin. The actual origin can be determined by fitting the profile along the $\phi$ axis, obtained from Figure~\ref{fig:dataPhiBP}~(a), to $p_0 + p_1\times \sin(\phi +p_2)$, where $p_N$,~\emph{N}=0-2 are the fit parameters. The $\phi$ profile and the fit are shown in Figure~\ref{fig:dataPhiBP}~(b).

The fit gives the center of the beam pipe to be ($-0.22\pm 0.04,-2.01 \pm 0.04$)~mm (the uncertainties quoted here are from the fit, and do not include any systematic effects). A shift of this magnitude is not unexpected, given the mechanical tolerances involved in placing the beam pipe along the center-line of the detector. Using the same procedure, the first and second layers of the pixel detector are found to be centered around ($-0.36\pm 0.03,-0.51 \pm 0.03$)~mm, and the third layer is centered around ($-0.19\pm 0.02,-0.29 \pm 0.02$)~mm.

\subsection{Details of modules in the pixel detector}
\label{pixels}

A module in the pixel detector has a complex structure~\cite{detector2}, and to explore finer details vertex positions are transformed from the global to the local pixel module coordinate system and all modules within one layer are overlaid. The excellent resolution of this technique makes such a detailed study of the inner detector elements possible. Transforming to the local frame also accounts for misalignments at the module level. In this coordinate system, the \emph{x(y)} axis is along the shorter (longer) edge of a module, and the \emph{z} axis points out of the plane of the module.

\begin{figure}[htbp]
\centering
\mbox{\subfigure[Data]{\includegraphics[width=0.5\textwidth]{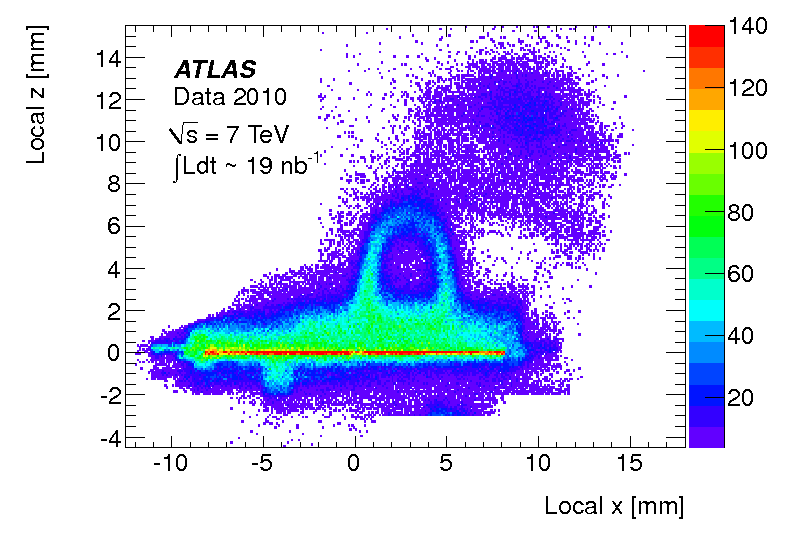}}\quad
\subfigure[MC]{\includegraphics[width=0.5\textwidth]{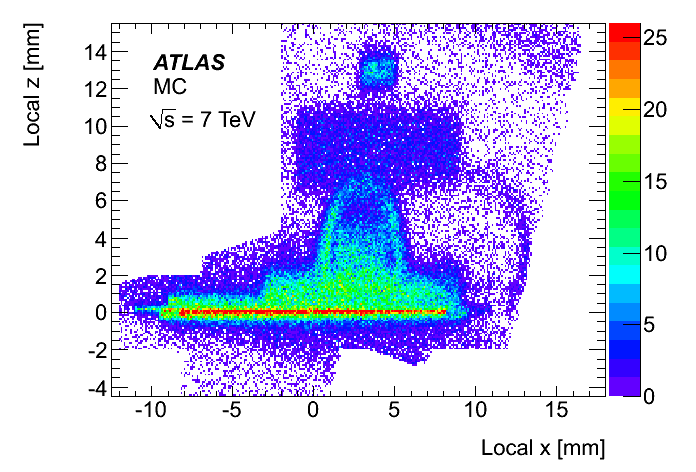}}}
\caption{(a) Data, (b) Non-diffractive MC. \emph{z} vs. \emph{x} in the local coordinate system for the first pixel detector layer, where $|z|$ in the global coordinate system is required to be less than 300~mm. The bin width is 0.1~mm in \emph{z} and \emph{x}. The limits on the minimum and maximum number of entries per bin have been
adjusted for presentational purposes, and to highlight various features.
\label{fig:dataMCLocalZX_Pix1_zoom}}
\end{figure}

A `zoomed-in' view of \emph{z} vs. \emph{x} coordinates for the first layer of the pixel detector in data and MC can be seen in Figures~\ref{fig:dataMCLocalZX_Pix1_zoom}~(a) and (b), respectively. The circular feature in both distributions at $0< x < 5$~mm is the cooling pipe, the rectangular features in the MC at $-1<x<9$~mm and $6<z<11$~mm and $2<x<5$~mm and $z\sim$ 12~mm are cables and connectors, which in reality are spread over a wider region, and also are in a slightly different location. The region with the most vertices, $-10<x<8$~mm and $z\sim 0$~mm, is the silicon sensor itself. The rectangular feature at $x\sim$ --5~mm and $z\sim$ --1~mm in the data plot shows the location of a capacitor. Also, there appears to be a difference in the density of vertices inside the cooling pipe; this is related to the fact that in the MC the cooling medium is in liquid phase, whereas in reality it is mainly in gaseous phase.

\section{Systematic uncertainties}
\label{systematics}

\subsection{Tracking efficiency}
\label{track_syst}

Previous studies have shown that the overall scale of the track reconstruction efficiency of charged particles in data is well simulated in the MC, and the main source of systematic uncertainties in the reconstruction efficiency of charged hadrons is the uncertain knowledge of the material in the inner detector~\cite{Syst1}. An increase (decrease) in material leads to an increase (decrease) in the number of hadronic interactions, hence to a decrease (increase) in the reconstruction efficiency. The effect on the number of reconstructed interaction vertices is a non-trivial interplay between these two effects. Since the overall scale of the reconstruction efficiency is well understood, an incorrect description of the inner detector material in the MC will lead to differences in the reconstruction efficiency as a function of the location of the vertex where the secondary tracks originate.

In order to make a comparison of the efficiency in data and MC as a function of vertex positions, independent of the estimated hadronic interaction rate, $\kshort$ decays are used. The momentum spectrum of $\kshort$ candidates in data agrees with MC, thereby making this technique feasible. They provide an ideal source of charged pions to study this issue. Pions produced in these decays have large values of transverse impact parameter (which, on average, increase with $\kshort$ decay distance), and probe the material in a similar manner as the tracks emerging from a secondary interaction vertex.
The $\kshort$ candidate mass distributions at various decay lengths\footnote{The choice of these regions is mainly driven by the need to maximize the sample size at large decay distances.}, viz., 39-45~mm, 46-60~mm, 60-85~mm, and above 85~mm, are fitted by a sum of two Gaussian functions (with common mean) for the signal and a first-order polynomial for the background. The $\kshort$ yield in each of these radial regions is normalized to the $\kshort$ yield inside the beam pipe (decay length within 10-25~mm), thus giving a radially-dependent ratio of yields. These ratios are determined separately in data and MC. For each radial region a double ratio using the ratio of yields in data and MC is determined.

If the dependence of the reconstruction efficiency for secondary tracks on vertex position was perfectly simulated in the MC, this double ratio would be unity. Although close to unity in most of the regions, this double ratio has the largest deviation in the 60-85~mm region, where it has a value of $0.93 \pm 0.02$. Consequently, this (largest) deviation from unity is taken as the systematic uncertainty on the efficiency of reconstructing secondary tracks, which corresponds to an uncertainty of 3.5\% per track. This is taken to be fixed over all allowed values of $\pt$, $\eta$, $\phi$, impact parameters of the track, and \emph{R} and \emph{z} of the originating vertex.

\subsection{Selection criteria during vertex finding}

Systematic effects in the vertex finding and fitting algorithms due to mismodelling of track parameters are determined by varying criteria to (a) merge nearby vertices, (b) uniquely assign tracks to a single vertex, and, (c) change the allowed range of $\chi^2$ for two-track vertices. In each case, values for selection criteria are (individually) varied in MC and data, and the difference between MC predictions and what was actually observed in data is found to be less than 1~\%. A total systematic uncertainty of 1\% is assigned due to these sources.

\subsection{Other sources}

To allow data to be compared to only non-diffractive MC, the contamination from single- and double-diffractive events in data is reduced by requiring at least 11 tracks at the primary vertex. MC studies suggest that this criterion still leaves a small amount of contamination of diffractive events ($\sim 1\%$). To investigate this, a stricter requirement is made on the track multiplicity at the primary vertex, expected to reduce the contamination from $\sim 1\%$ to $\sim 0.1\%$.  This tighter requirement should have no effect on the non-diffractive MC other than to reduce the overall statistics. In data, the change in yield is $\sim 0.6\%$ more than the expectation from the non-diffractive MC sample, and this difference is taken to be the systematic uncertainty due to this source.

When comparing yields in data and MC, the latter is corrected because it has fewer primary tracks. This was discussed in Section~\ref{trackMC}. Different criteria are used to decide what constitutes a primary track, e.g., varying selections on transverse and longitudinal impact parameters, number of hits in the pixel detector. Correction factors for primary and secondary tracks are investigated separately (the latter selected by requiring the transverse impact parameter relative to the primary vertex to be larger than 5~mm), and an average is determined based on estimates (from MC) for the fraction of interactions that are due to secondary tracks. A systematic uncertainty of $1 \%$ is assigned from this source. This procedure does not explicitly account for neutral hadrons, viz., neutrons and neutral kaons. However, the number of neutrons produced during fragmentation is related to the proton yield due to isospin invariance in strong interactions; similarly, the yield of neutral kaons is related to the yield of charged kaons. In addition, many of the $\kshort$'s will decay before they can interact. Finally, the yield of primary neutrons and kaons is about one-quarter of the yield of primary protons and charged pions~\cite{pythia6}. Since the procedure corrects for charged hadrons, i.e., protons and kaons, the systematic uncertainty from not explicitly accounting for neutral hadrons is expected to be small and is neglected.

To account for the shift in the location of the beam pipe, yields are determined after correcting for its position offset. Varying the \emph{x,y} offsets within $\pm 1 \sigma$ of the central values leads to changes in yields of less than 0.1\%. In the case of the pixel detector, the effect of correcting the position offset on the yields is equally small. Hence, no systematic uncertainty is assigned.

\subsection{Total systematic uncertainty}

The systematic uncertainty on track reconstruction is propagated into the total uncertainty by using MC and randomly removing 3.5\% of the tracks from true nuclear interaction vertices that match reconstructed tracks passing all other selection criteria, and observing that 6.3\% of the true vertices are lost. This decrease is taken to be the systematic uncertainty on the ratio of vertex yields in data and MC from this source. Combining this with all other sources leads to a total systematic uncertainty of 6.6\% on the ratios.

An additional uncertainty arises from the modeling of hadronic interactions in {\sc geant4}, but this is hard to quantify. This motivates the study presented in Section 7.2, which investigates vertex yields in the beryllium part of the beam pipe. The good agreement between the vertex yield measured in data and simulation when focusing on the well-known areas of the detector gives confidence in the modeling. The uncertainty from the modeling of the composition of primary particles in {\sc pythia6} is expected to be much smaller.

\section{Numerical comparison of data and MC}
\label{compareSec}

\subsection{Vertex yields}
\label{compare}

Table~\ref{tab:DataYields} displays yields in selected material layers. The chosen regions include the silicon sensors, as well as supports, cables and services. The beam pipe envelope includes an $800~\mu$m thick beryllium cylinder, 4~mm of aerogel and thin layers of materials such as kapton tape and coatings. The yields in data presented here and later are given after the position offset correction procedure described in Section~\ref{beampipePos}. Given that most of the vertices lie in $|z|<300$~mm, all the following figures and results will be restricted to this region. Also, MC studies indicate that the purity of reconstructed vertices (i.e. the fraction of all reconstructed vertices which match a true MC vertex) degrades at larger values of $|z|$. In this restricted \emph{z} region, and for radii at or greater than the beam pipe, the data sample consists of more than $10^6$ vertices, with about 42\% of them containing two oppositely charged tracks, 51\% having two tracks of the same charge, and the remaining 7\% having three or more tracks, which is in good agreement with MC, where the corresponding fractions are 45\%, 49\% and 6\%, respectively.

The yield of reconstructed vertices is a function of radius. Some of the reasons for this dependence are (a) a decrease in the reconstruction efficiency of secondary tracks emerging from hadronic interactions as a function of radius, (b) a decrease in the number of primary particles that intersect successive layers within $|z| < 300$~mm, thereby leading to fewer interactions, and, (c) the difference in the amount of material in various layers. In addition, the momentum spectrum of tracks intersecting the outer layers is slightly softer on average than those intersecting the inner layers, and this may also contribute to a difference in the number of interactions. The efficiency to find reconstructible vertices, i.e. vertices that have at least two tracks with $\pt$ and $\eta$ satisfying the selection criteria, ranges from about 5\% at the beam pipe to 1\% at the third layer of the pixel detector, and 0.5\% at the first SCT layer; these include efficiencies for both track and vertex reconstruction steps.

\begin{table}[htbp]
  \begin{center}
    \caption{Yield of reconstructed vertices (data).
    \label{tab:DataYields}}
    \begin{tabular}{|l|c|c|}\hline
      Vertex Radius range&\multicolumn{1}{|c|}{Yield $|z|<300$~mm}&\multicolumn{1}{|c|}{Yield $|z|>300$~mm}\\
      \hline \hline
      Beam Pipe (28-36~mm)       & 542643& 1040\\
      1$^{st}$ pixel layer (47-72~mm) & 517835 & 3541 \\
      2$^{nd}$ pixel layer (85-110~mm) & 133395  & 6611\\
      3$^{rd}$ pixel layer (119-145~mm)& 83443  & 24960\\ \hline
      1$^{st}$ SCT layer (275-320~mm) & 9746  & 11373\\ \hline
    \end{tabular}
  \end{center}
\end{table}

\subsection{Details of interactions in beryllium part of beam pipe}
\label{kinematics}

To address the additional uncertainty arising from the modeling of hadronic processes in {\sc geant4}~\cite{geantref,Wright}, vertices are reconstructed in the beryllium part of the beam pipe. The yields of such vertices and the kinematic variables describing them are compared between data and MC. This region is chosen because the material is a single element and its dimensions (radial thickness of $800~\mu$m) and location are precisely known. Since the beam pipe is close to the collision point, the secondary track reconstruction efficiency and the purity of the reconstructed vertices are high. As discussed in Section~\ref{track_syst}, secondary track reconstruction is well simulated in MC. Under the assumption that the composition of primary particle types in data is correctly predicted in {\sc pythia6}~\cite{pythia6}, such comparisons allow to check the quality of the modeling in {\sc geant4}. According to the MC, particles interacting in the beam pipe are expected to be charged pions (62\%), protons (17\%), neutrons (14\%) and kaons (7\%). The momenta of charged particles that impinge on the beam pipe agree between data and MC, as shown in Figure~\ref{fig:compareDataMCPrimTrkBP}. The observed vertex yields in MC and data are 70297 and 227921, respectively, and the rate of secondary vertices per event is $(5.57 \pm 0.02)\times 10^{-3}$ in the MC, and $(5.58 \pm 0.01)\times 10^{-3}$ in data, resulting in a ratio of yields in data to MC of $1.002 \pm 0.004 \pm 0.066$, where the first uncertainty is  statistical and the second is systematic. For determining the yield in data, the radial position of the vertices are taken relative to (--0.22,--2.0)~mm. These rates include fake vertices. From MC studies, the purity of the reconstructed vertices is estimated to be $\sim$ 82\%.

\begin{table}[htbp]
  \begin{center}
    \caption{Comparison of track multiplicity, in MC and data, at secondary vertices in the beryllium part of the beam pipe (statistical uncertainties only).
    \label{tab:BeNtrks}}
    \begin{tabular}{|l|c|c|}\hline
      Track Multiplicity &\multicolumn{1}{|c|}{MC (stat.)}&\multicolumn{1}{|c|}{Data (stat.)}\\
      \hline \hline
      Fraction of 2-track vertices & $0.916 \pm 0.001$ & $0.906 \pm 0.001$ \\
      Fraction of 3-track vertices & $0.079 \pm 0.001$ & $0.087 \pm 0.001$ \\
      Fraction of $\geq$ 4-track vertices & $0.0058 \pm 0.0003$& $0.0076 \pm $ 0.0002\\ \hline
    \end{tabular}
  \end{center}
\end{table}

\begin{figure}[htbp]
\begin{center}
\centering
\mbox{\subfigure[Invariant Mass]{\includegraphics[width=0.5\textwidth]{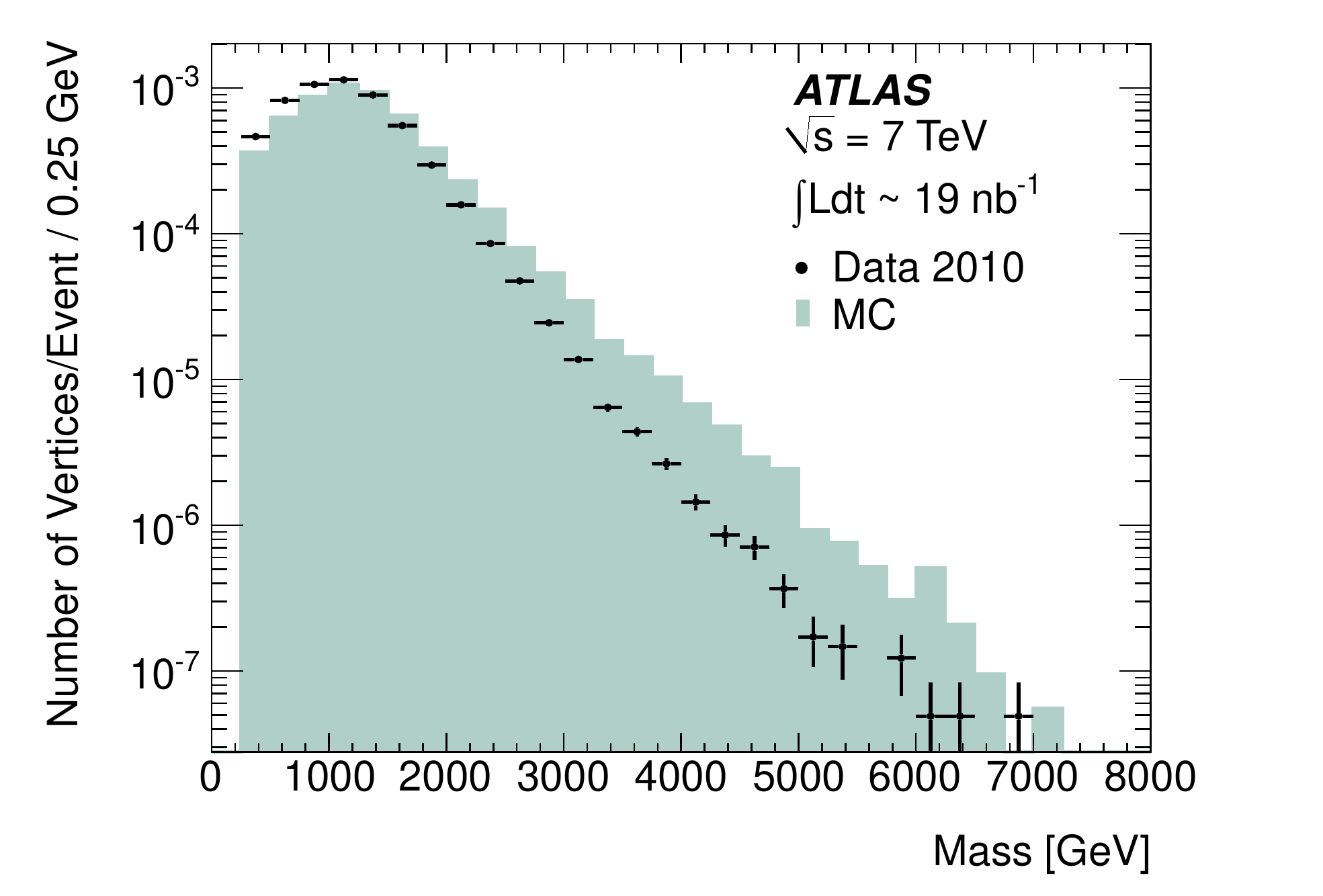}}\quad
\subfigure[$|\sum \vec{p}\;|$]{\includegraphics[width=0.5\textwidth]{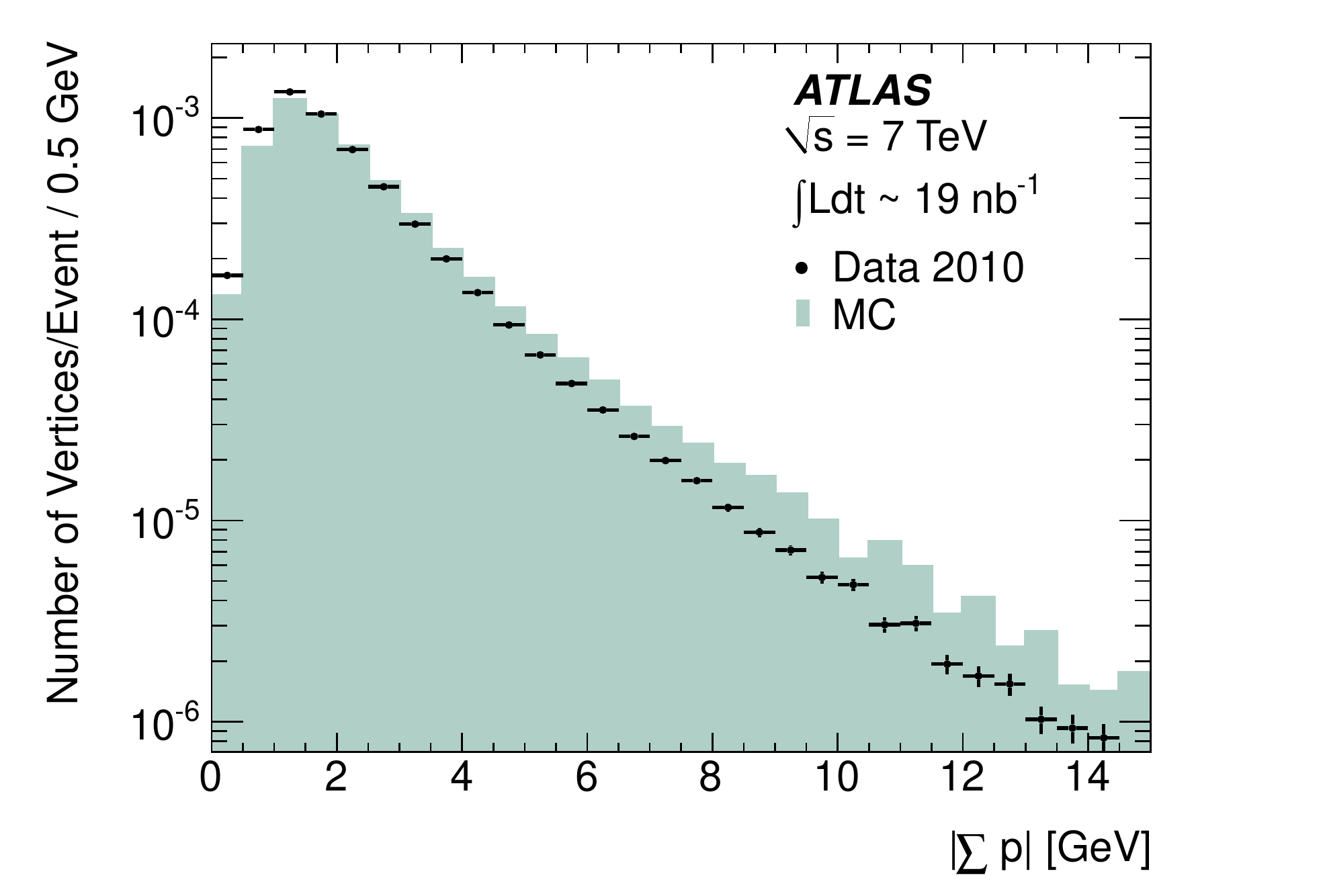}}}
\caption{Kinematic variables for reconstructed vertices in the beryllium part of the beam pipe for data (points) and MC (filled histogram).
\label{fig:BedataMCcompareVars}}
\end{center}
\end{figure}

The breakdown of track multiplicity for the reconstructed vertices in MC and data are shown in Table~\ref{tab:BeNtrks}. Although the hierarchy of track multiplicity is well reproduced, there appears to be some numerical discrepancies, which cannot be explained only by the systematic uncertainty on the reconstruction efficiency of secondary tracks, and may point to insufficient accuracy in the modeling of hadronic interactions in {\sc geant4}. Figure~\ref{fig:BedataMCcompareVars} presents distributions of the invariant mass of the vertex (left), and $|\sum \vec{p}\;|$ (right), i.e., the magnitude of the vector sum of the momenta of the outgoing tracks. These variables are useful in understanding the kinematics of the interaction. Although the overall shapes of the mass and $|\sum \vec{p}\;|$ distributions are in reasonable agreement between data and MC, there are some differences, e.g., in the high mass tail, which could be used to further refine models of hadronic interactions.

\subsection{Comparison of vertex yields in data and MC}
\label{yield}

Table~\ref{tab:compareYields} presents a comparison of the rate of interaction vertices/event between data and MC. Only statistical uncertainties are quoted for the yields for both data and MC, while the ratio of yields also includes the systematic uncertainty. These rates include fake vertices. From MC studies, the purity of reconstructed vertices in this restricted \emph{z} region is estimated to be $\sim$ 82\%, 73\%, 78\%, 46\%, and 49\%, respectively, at the five material layers listed in the table\footnote{To determine if the reconstructed vertex position matches that of the a true interaction vertex, the \emph{R} and \emph{z} position of the two are required to be within ranges that depend on the vertex resolutions discussed in Section~\ref{vtxres}.}. In general, the agreement is very good, i.e., at the ~7\% level of the systematic uncertainty. The beam pipe envelope contains beryllium, layers of aerogel, kapton tape and coatings, and the pixel detector and SCT regions include supports, cables, and services.

\begin{table}[htbp]
  \begin{center}
    \caption{Comparison of rates of reconstructed vertices per event in MC and data. The pixel and SCT layers include detector modules, services and support structures. Statistical and systematic uncertainties are listed.
    \label{tab:compareYields}}
    \begin{tabular}{|l|c|c|c|}\hline
      Vertex Radius range &\multicolumn{1}{|c|}{MC($\times 10^{-3}$)}&\multicolumn{1}{|c|}{Data($\times 10^{-3}$)} &\multicolumn{1}{|c|}{Data/MC} \\
                           &   (stat.)        & (stat.)   & (stat., syst.) \\\hline \hline
      Beam Pipe (28-36~mm) & $12.76 \pm 0.03 $& $13.27 \pm 0.02 $ & $1.040\pm 0.003 \pm 0.069$\\
      1$^{st}$ pixel layer (47-72~mm)& $ 13.40 \pm 0.03$ & $12.66 \pm 0.02 $ & $0.945\pm 0.003 \pm 0.063$\\
      2$^{nd}$ pixel layer (85-110~mm)&  $3.47 \pm 0.02$ & $ 3.26 \pm 0.01$ & $0.94\pm 0.01\pm 0.06$\\
      3$^{rd}$ pixel layer (119-145~mm)& $ 1.97 \pm 0.01$  & $ 2.04 \pm 0.01$ & $1.04\pm 0.01 \pm 0.07$\\
      1$^{st}$ SCT layer (275-320~mm)&  $0.22 \pm 0.004$ & $ 0.24 \pm 0.002$ & $1.09\pm 0.03 \pm 0.07$\\ \hline
    \end{tabular}
  \end{center}
\end{table}

\subsection{Vertex yields within the modules of the pixel detector}
\label{numerical_pixels}

To make quantitative comparisons of the material in the pixel detector modules, yields of found vertices are compared in regions that lie within the bulk of the module and supporting structure. These results are presented in Table~\ref{tab:pixelDW}. Vertices within a box whose corners in (\emph{x,z}), as measured in the local module coordinate system, are at (--11.5,--2.1)~mm and (11.1,1.1)~mm, are counted. Only statistical uncertainties are quoted for the yields in both data and MC, while the ratio of yields includes the systematic uncertainty. The agreement is again very good. Although all the modules are the same, the systematic uncertainty is not completely correlated between the three layers. For instance, vertex resolutions worsen with increasing radius, leading to a larger migration of vertices in and out of the chosen boxes.

In addition, the number of vertices inside the cooling pipe is different in data and MC, and results for the first layer, where the effect is most clearly visible, indicate that the fraction of vertices inside the cooling pipe relative to all vertices in that layer is $ 4.6 \pm 0.1 \%$ in MC and $1.7 \pm 0.1 \%$ in data, where the uncertainties are statistical. Since this is a ratio, systematic uncertainties largely cancel.

\begin{table}[htbp]
  \begin{center}
    \caption{Comparison of rates of reconstructed vertices per event in MC and data in the pixel detector modules.
    \label{tab:pixelDW}}
    \begin{tabular}{|l|c|c|c|}\hline
      Vertex Location&\multicolumn{1}{|c|}{MC($\times 10^{-3}$)}&\multicolumn{1}{|c|}{Data($\times 10^{-3}$)}&\multicolumn{1}{|c|}{Data/MC}\\
                      & (stat.)          & (stat.)          & (stat., syst.) \\\hline \hline
      1st pixel layer & $ 6.08 \pm 0.02$ & $6.16 \pm 0.01 $ & $1.01\pm 0.01 \pm 0.07 $\\
      2nd pixel layer &  $1.65 \pm 0.01$ & $1.71 \pm 0.01 $ & $1.04\pm 0.01 \pm 0.07 $\\
      3rd pixel layer & $1.40 \pm 0.01$  & $1.43 \pm 0.01 $ & $1.02\pm 0.01 \pm 0.07 $ \\ \hline

    \end{tabular}
  \end{center}
\end{table}

\section{Conclusions}

Secondary vertices due to hadronic interactions of primary particles have been reconstructed and used to study the distribution of material within the ATLAS inner detector volume. Reconstruction of secondary vertices far from the primary vertex uses a subset of tracks that are not normally used in most analyses, thus, this analysis provides an interesting challenge for tracking algorithms optimized for tracks coming from the primary vertex.

The reconstructed secondary vertices have excellent spatial resolution, approximately 0.2-1~mm, in both longitudinal and transverse directions. This resolution is significantly better than the spatial resolution of vertices produced by photon conversions, which are routinely used for material estimation. This leads to a precise radiography of the as-built tracking sub-systems and facilitates comparison with the implementation of the detector geometry in MC. For instance, the detailed structure of modules in the pixel detector has been investigated, and the distribution of material in data and MC are found to be in very good agreement. However, some discrepancies in the MC model have been discovered, the most important being that, in reality, the beam pipe is not centered around the (0,0) position. Another discrepancy is in the density of the fluid used to cool the pixel modules. In reality, the fluid is a mix of liquid and gaseous phases, whereas in MC, it is assumed to be a liquid. These features have been included in newer versions of the MC.

The estimation of the exact amount of material based on the number of reconstructed vertices is affected by many sources of systematic uncertainty, viz., the secondary track and vertex reconstruction efficiencies, the composition of primary particles that interact in the inner detector, their $\pt, \eta$ distributions, and the accuracy of hadronic interaction modeling in {\sc geant4}. In the current analysis the experimental systematic uncertainties, i.e., those arising from track and vertex reconstruction, have been estimated from data. Differences between data and MC in the $\pt$ and $\eta$ distributions of the primary tracks are accounted for via a reweighting procedure. This leads to an estimate of the total experimental systematic uncertainty of about 7\%.  Results obtained for the well-known parts of the inner detector, the beam pipe and pixel modules, provide confirmation for this estimate.

There are additional sources of systematic uncertainty, and the numerical estimate of those uncertainties is outside the scope of the current analysis. They are due to an incomplete knowledge of the composition of primary particles in non-diffractive events, in particular, the flux of neutral particles, and the modeling of hadronic interactions in {\sc geant4}. The primary particle composition in {\sc pythia6} is based on data from previous experiments, e.g., those at the Large Electron-Positron Collider at CERN. Due to the nature of the fragmentation process, it is expected that the total number of primary particles in \emph{pp} collisions is more than at $e^+ e^-$ collisions, but the fractions of the pions, kaons, baryons, etc., are similar. However, this needs to be verified. In addition, the current study demonstrates that, in general, the quality of {\sc geant4} predictions is quite good, but some differences in the distributions of kinematic variables do exist between data and MC, and these should be taken into account in future refinements.

\section{Acknowledgements}

We thank CERN for the very successful operation of the LHC, as well as the
support staff from our institutions without whom ATLAS could not be
operated efficiently.

We acknowledge the support of ANPCyT, Argentina; YerPhI, Armenia; ARC,
Australia; BMWF, Austria; ANAS, Azerbaijan; SSTC, Belarus; CNPq and FAPESP,
Brazil; NSERC, NRC and CFI, Canada; CERN; CONICYT, Chile; CAS, MOST and
NSFC, China; COLCIENCIAS, Colombia; MSMT CR, MPO CR and VSC CR, Czech
Republic; DNRF, DNSRC and Lundbeck Foundation, Denmark; ARTEMIS, European
Union; IN2P3-CNRS, CEA-DSM/IRFU, France; GNAS, Georgia; BMBF, DFG, HGF, MPG
and AvH Foundation, Germany; GSRT, Greece; ISF, MINERVA, GIF, DIP and
Benoziyo Center, Israel; INFN, Italy; MEXT and JSPS, Japan; CNRST, Morocco;
FOM and NWO, Netherlands; RCN, Norway; MNiSW, Poland; GRICES and FCT,
Portugal; MERYS (MECTS), Romania; MES of Russia and ROSATOM, Russian
Federation; JINR; MSTD, Serbia; MSSR, Slovakia; ARRS and MVZT, Slovenia;
DST/NRF, South Africa; MICINN, Spain; SRC and Wallenberg Foundation,
Sweden; SER, SNSF and Cantons of Bern and Geneva, Switzerland; NSC, Taiwan;
TAEK, Turkey; STFC, the Royal Society and Leverhulme Trust, United Kingdom;
DOE and NSF, United States of America.

The crucial computing support from all WLCG partners is acknowledged
gratefully, in particular from CERN and the ATLAS Tier-1 facilities at
TRIUMF (Canada), NDGF (Denmark, Norway, Sweden), CC-IN2P3 (France),
KIT/GridKA (Germany), INFN-CNAF (Italy), NL-T1 (Netherlands), PIC (Spain),
ASGC (Taiwan), RAL (UK) and BNL (USA) and in the Tier-2 facilities
worldwide.

\clearpage
\input{atlas_authlist}

\end{document}

%% file: atlas_authlist.tex
\begin{flushleft}
{\Large The ATLAS Collaboration}

\bigskip

G.~Aad$^{\rm 48}$,
B.~Abbott$^{\rm 111}$,
J.~Abdallah$^{\rm 11}$,
A.A.~Abdelalim$^{\rm 49}$,
A.~Abdesselam$^{\rm 118}$,
O.~Abdinov$^{\rm 10}$,
B.~Abi$^{\rm 112}$,
M.~Abolins$^{\rm 88}$,
H.~Abramowicz$^{\rm 153}$,
H.~Abreu$^{\rm 115}$
% ...
\bigskip

% ...
$^{10}$ Institute of Physics, Azerbaijan Academy of Sciences, Baku, Azerbaijan\\
$^{11}$ Institut de F\'isica d'Altes Energies and Departament de F\'isica de la Universitat Aut\`onoma  de Barcelona and ICREA, Barcelona, Spain\\
% ...
$^{48}$ Fakult\"{a}t f\"{u}r Mathematik und Physik, Albert-Ludwigs-Universit\"{a}t, Freiburg i.Br., Germany\\
$^{49}$ Section de Physique, Universit\'e de Gen\`eve, Geneva, Switzerland\\
% ...
$^{88}$ Department of Physics and Astronomy, Michigan State University, East Lansing MI, United States of America\\
% ...
$^{111}$ Homer L. Dodge Department of Physics and Astronomy, University of Oklahoma, Norman OK, United States of America\\
$^{112}$ Department of Physics, Oklahoma State University, Stillwater OK, United States of America\\
% ...
$^{115}$ LAL, Univ. Paris-Sud and CNRS/IN2P3, Orsay, France\\
% ...
$^{118}$ Department of Physics, Oxford University, Oxford, United Kingdom\\
% ...
$^{153}$ Raymond and Beverly Sackler School of Physics and Astronomy, Tel Aviv University, Tel Aviv, Israel\\
% ...
\end{flushleft}
